\documentclass[]{iopart}
\usepackage{iopams, graphicx, hyperref}
\usepackage[usenames,dvipsnames]{color}
\usepackage[normalem]{ulem}

\newcommand{\half}{{\tfrac{1}{2}}}

\newcommand{\hateq}{\; \hat{=} \;}

\newcommand{\tfrac}[2]{\textstyle \frac{#1}{#2}}
\newcommand{\be}{\mathbf{e}}
\newcommand{\Hobc}[1]{${\mathcal B}_{#1}$}

\begin{document}

\title[Implementation of absorbing boundary conditions 
       for the Einstein equations]
  {Implementation of higher-order absorbing boundary conditions 
   for the Einstein equations}

\author{Oliver Rinne$^{1,2,3}$, Luisa T.~Buchman$^{3,4}$, 
  Mark A.~Scheel$^{3}$ and Harald P.~Pfeiffer$^{3}$}

\address{
  $^{1}$Department of Applied Mathematics and Theoretical
  Physics, Centre for Mathematical Sciences, Wilberforce Road,
  Cambridge CB3 0WA, UK\\
  $^{2}$King's College, Cambridge CB2 1ST, UK\\
  $^{3}$Theoretical Astrophysics 130-33, California Institute of
  Technology, Pasadena, CA 91125, USA\\
  $^{4}$Center for Relativity, University of Texas at Austin, 
  Austin, TX 78712, USA
}

\begin{abstract}
  We present an implementation of absorbing boundary conditions for
  the Einstein equations based on the recent work of Buchman and
  Sarbach.  In this paper, we assume that spacetime may be
  linearized about Minkowski space close to the outer boundary, which
  is taken to be a coordinate sphere.  We reformulate the boundary
  conditions as conditions on the gauge-invariant
  Regge-Wheeler-Zerilli scalars.  Higher-order radial derivatives are
  eliminated by rewriting the boundary conditions as a system of ODEs
  for a set of auxiliary variables intrinsic to the boundary.  From
  these we construct boundary data for a set of well-posed
  constraint-preserving boundary conditions for the Einstein equations
  in a first-order generalized harmonic formulation.  This
  construction has direct applications to outer boundary conditions in
  simulations of isolated systems (e.g., binary black holes) as well
  as to the problem of Cauchy-perturbative matching.  As a test
  problem for our numerical implementation, we consider linearized
  multipolar gravitational waves in TT gauge, with angular momentum
  numbers $\ell=2$ (Teukolsky waves), $3$ and $4$.  We demonstrate
  that the perfectly absorbing boundary condition \Hobc{L} of
  order $L=\ell$ yields no spurious reflections to linear order in
  perturbation theory.  This is in contrast to the lower-order
  absorbing boundary conditions \Hobc{L} with $L<\ell$, which
  include the widely used freezing-$\Psi_0$ boundary condition that
  imposes the vanishing of the Newman-Penrose scalar $\Psi_0$.
\end{abstract}



\section{Introduction}

Many situations of astrophysical interest can be described to good
approximation as an \emph{isolated system}: an asymptotically flat
spacetime containing a compact self-gravitating source.  The study of
such systems requires the solution of Einstein's equations on an
unbounded domain.  One of the major problems in numerical relativity
is how to accomplish this with finite computer resources.  The most
common approach is based on the Cauchy formulation of general
relativity.  Here the spatial computational domain is truncated at a
finite distance and boundary conditions (BCs) are imposed at this
artificial timelike boundary. 
These BCs must satisfy a number of requirements, the most important 
ones being that (i) the BCs must be compatible with the constraint 
equations that hold within the spatial slices, i.e.~the BCs must be 
\emph{constraint preserving}, (ii) the resulting initial-boundary 
value problem (IBVP) must be well posed, and (iii) the BCs should minimize
spurious reflections of gravitational radiation, i.e.~the BCs should
be \emph{absorbing}. (We call BCs that completely eliminate such reflections 
\emph{perfectly absorbing}.)  
It is important here to distinguish between spurious
reflections on the one hand, and non-spurious reflections such as
the backscatter off a curved background spacetime on the other hand.
Perfectly absorbing BCs are defined to be exactly satisfied by the general
retarded solution at the outer boundary, i.e.~they only eliminate
the spurious reflections but preserve non-spurious reflections such 
as backscatter.
While considerable progress has been made on the first two requirements
above, i.e.~constraint preservation and well posedness,
the third one on the absorbing properties of the BCs
has not been addressed to the same extent.  
It is however of prime importance if accurate approximations to the
gravitational radiation emitted by the source are to be computed,
as required for instance by gravitational wave data analysis.

Theoretical progress on the construction of absorbing BCs for the
Einstein field equations was made by Buchman and Sarbach in
\cite{Buchman2006, Buchman2007}.  A hierarchy of local BCs
\Hobc{L} was proposed that is perfectly absorbing for all radiation with
angular momentum numbers $\ell \leqslant L$, where $L$ is an arbitrary given 
number. These BCs were obtained by
studying solutions to the Bianchi equations describing linearized
gravitational waves and finding a condition on the outgoing solutions
that was satisfied exactly. It turns out that these BCs are a
generalization to the Einstein equations of the well-known
Bayliss-Turkel BCs \cite{Bayliss1980} for the scalar wave
equation.  Initially \cite{Buchman2006}, the BCs were formulated for a
flat background spacetime; first-order corrections dealing with the
spacetime curvature and the backscatter on a Schwarzschild black hole
background were included in \cite{Buchman2007}.

The objective of the present paper is to reformulate the
Buchman-Sarbach BCs \Hobc{L} in such a way that they can be
incorporated into a full set of constraint-preserving BCs for the
Einstein equations, and to implement and test such BCs.  Before
describing our approach in more detail, we mention a number of
previous works on and alternatives to absorbing BCs.

In \cite{Novak2004}, the Bayliss-Turkel BCs were implemented up to
quadrupolar order ($\ell=2$) for the scalar wave equation.  The
reformulation of the BCs and the numerical method used in that paper
are similar to ours, but we note that we treat gravitational rather
than scalar waves.  A different approach to absorbing BCs is based on
fast-converging series expansions of \emph{exact nonlocal} BCs
\cite{Alpert2000}.  This was applied to the construction of exact
absorbing BCs for the Regge-Wheeler and Zerilli (RWZ) equations in a
series of papers by Lau \cite{Lau2004,Lau2004a,Lau2005}. The RWZ
equations describe linear gravitational perturbations about a
Schwarzschild black hole, and they play a central role in our approach
as well. 
A general framework for the construction of absorbing
boundaries is Cauchy-perturbative matching (CPM)
\cite{Abrahams1998,Rupright1998,Rezzolla1999,Zink2006}.  Here one
matches solutions of the Einstein equations in the interior of a
compact domain to solutions of the linearized equations in the
exterior, represented by, for example, the RWZ equations.  As we shall
point out below, this approach is again closely related to ours.  In
addition, instead of matching to solutions of linearized gravity, one
can match to an outer module solving the Einstein equations in the
characteristic formulation, which is particularly well suited to the
extraction of gravitational radiation (see \cite{Winicour2005} for a
review article).  Finally, a very promising method consists in solving
the Einstein equations on hyperboloidal slices that are everywhere
spacelike but that asymptotically approach null infinity (see
\cite{FrauendienerLRR} for a review article).

Our approach is based on the Buchman-Sarbach BCs, and in the following
we describe the main idea of our algorithm and the organization of
this paper in more detail.  We assume that close to the outer
boundary, the spacetime metric can be described by linear
perturbations of Minkowski spacetime in the standard coordinates.
This is usually a good approximation if the outer boundary is placed
sufficiently far out (we intend to generalize our work to a
Schwarzschild background in the future).  
Note that on a flat background (unlike on a curved background
where there is backscatter), the general retarded solution
is purely outgoing, so in this case perfectly absorbing BCs should 
eliminate precisely the ingoing solution. 
We furthermore assume that
the outer boundary is a sphere of constant coordinate radius.  The
original Buchman-Sarbach conditions involve higher-order radial
derivatives of the Newman-Penrose scalar $\Psi_0$, making it a
non-trivial task to implement them numerically.  Our strategy is to
work with the RWZ scalars instead of $\Psi_0$ because on a flat-space
background, the Buchman-Sarbach BCs on the RWZ scalars are precisely
the same as the Bayliss-Turkel BCs for the scalar wave equation
(section \ref{s:RWZBCs}). Additionally, since the RWZ scalars obey
closed wavelike evolution equations (the RWZ equations), we can draw
on previous work in computational mathematics that successfully
implements higher-order absorbing BCs for the scalar
wave equation. Following \cite{Hagstrom1998,Huan2000,Givoli2001}, we
eliminate the higher-order radial derivatives by rewriting the BCs as
a system of ordinary differential equations (ODEs) for a set of
auxiliary variables that need only be defined at the boundary (section
\ref{s:AuxVars}).  This auxiliary system is completely equivalent to
the absorbing BCs imposed on the RWZ scalars.

In the interior of the computational domain, the full (nonlinear)
Einstein equations are evolved.  Hence we must transfer information
between them and the auxiliary system at the boundary. The RWZ scalars
are computed from the spacetime metric in section \ref{s:Extraction}
following the gauge-invariant treatment of Sarbach and Tiglio
\cite{Sarbach2001}.  In turn, we must provide boundary data for the
incoming characteristic fields of the Einstein equations (section
\ref{s:EinsteinBCs}).  We use a first-order formulation of these
equations in generalized harmonic gauge (see \cite{Lindblom2006} and 
references therein). In \cite{Kreiss2006}, a
set of constraint-preserving and well-posed BCs in Sommerfeld form was
proposed for the harmonic Einstein equations.  These BCs contain
certain free boundary data, and only very special values of these data
will yield BCs that are also absorbing.  The novelty of our approach
lies in the construction of such absorbing boundary data from the
auxiliary system at the boundary (section \ref{s:Reconstruction}).  We
stress here that these data could equally well be obtained from
exterior solutions of the RWZ equations in the CPM approach, so that
our work can also be viewed as an explicit prescription for
constructing BCs for the generalized harmonic Einstein equations from such
perturbative solutions.

In summary, our algorithm consists in three basic steps:
(i) extraction of the RWZ scalars from the spacetime metric at the
boundary,
(ii) evolution of the auxiliary variables at the boundary,
and (iii) construction of absorbing boundary data for the Einstein equations
from the auxiliary variables.

We evolve the generalized harmonic Einstein equations using a pseudospectral 
collocation method, the Caltech-Cornell Spectral Einstein Code (SpEC). 
Some details of the numerical method relevant to the present work are
described in section \ref{s:NumEvMethod}.

As a first test problem of our implementation, we consider exact
solutions of linearized gravity, multipolar gravitational waves in 
transverse-traceless (TT) gauge (section \ref{s:MultipolarSoln}).
The quadrupolar ($\ell=2$) solution was first given in explicit form by
Teukolsky \cite{Teukolsky1982}, and it has recently been generalized 
to arbitrary $\ell$ in \cite{Rinne2008c}.
The BC \Hobc{L=\ell} is perfectly absorbing for these solutions. 
We set initial data for $\ell=2,3,4$ (section
\ref{s:InitialData}) and evolve them with this BC, \Hobc{L=\ell}, 
comparing the 
numerically extracted RWZ scalars with their analytical counterparts
(section \ref{s:MultipolarEvoln}).  We show that the difference
between the two decays (at least) quadratically with amplitude so that
there are no spurious reflections to linear order in perturbation
theory, as expected.  In contrast, \Hobc{L<\ell}
are shown to cause reflections at leading (linear) order.  
The $L=1$ case, \Hobc{1},
is equivalent to imposing the vanishing of the Newman-Penrose scalar
$\Psi_0$ at the outer boundary, a BC that is often used in numerical
relativity \cite{Bardeen2002,Kidder2005,Sarbach2005,Scheel2006,Rinne2007,
Boyle2007,Scheel2008} and that is referred to as the freezing-$\Psi_0$ BC.
There is a residual gauge freedom at the boundary in generalized 
harmonic gauge (section \ref{s:BdyOnGauge}), and we show that our 
numerical results are insensitive to the particular choice of gauge 
boundary data.  
We also compute approximate reflection coefficients from our numerical
evolutions and compare them with the theoretical predictions of
\cite{Buchman2006} (section \ref{s:ReflCoeff}).

We conclude and give an outlook onto future work in section \ref{s:Concl}.


\section{Formulation of the boundary conditions}
\label{s:Formulation}

In this section, we derive our reformulation of the higher-order absorbing
BCs \Hobc{L} proposed in \cite{Buchman2006,Buchman2007}.
We begin by showing how they can be expressed as BCs
on the RWZ scalars (section \ref{s:RWZBCs}).
Radial derivatives are eliminated by introducing a system of auxiliary 
ODEs at the boundary (section \ref{s:AuxVars}).
Next we explain how the RWZ scalars are extracted from the perturbed
spacetime metric, following the gauge-invariant treatment of 
Sarbach and Tiglio \cite{Sarbach2001} (section \ref{s:Extraction}).
Finally we describe the BCs that we impose on the
Einstein equations in a first-order generalized harmonic formulation of the
Einstein equations (section \ref{s:EinsteinBCs}).
Boundary data are constructed from the auxiliary system that
correspond precisely to the desired absorbing BC
(section \ref{s:Reconstruction}).


\subsection{Absorbing boundary conditions for the RWZ scalars}
\label{s:RWZBCs}

Although our approach is not limited to this case, 
we assume in this paper that spacetime near the outer
boundary can be described by linear perturbations of a flat
background spacetime in standard Minkowski coordinates. The
spacetime metric $g_{\alpha\beta}$ is written as
\begin{equation}
  g_{\alpha\beta} = \mathring{g}_{\alpha\beta} + \delta g_{\alpha\beta}.
\end{equation}
We assume that the background metric $\mathring{g}_{\alpha\beta}$ is a 
direct product 
\begin{equation}
  \mathring{g} = \tilde g_{ab} dx^a dx^b + r^2 \hat g_{AB} dx^A dx^B,
\end{equation}
where $\tilde g=-dt^2+dr^2$ is the standard Minkowski metric on a
2-manifold $\tilde M$ and $\hat g = d\theta^2 + \sin^2\theta \;
d\phi^2$ is the standard metric on the 2-sphere.  Throughout this
paper, Greek indices $\alpha,\beta,\ldots$ are spacetime indices,
lower-case Latin indices $a,b,\ldots$ range over $t$ and $r$, and
upper-case Latin indices $A,B,\ldots$ range over $\theta$ and $\phi$.
The covariant derivative compatible with the
metric $\tilde g$ ($\hat g$) will be denoted by $\tilde \nabla$ ($\hat
\nabla$) and the volume element by $\tilde \epsilon_{ab}$ ($\hat
\epsilon_{AB}$).
The covariant derivative $\tilde \nabla$ is sometimes also denoted by
a vertical bar ($|$).

The hierarchy of absorbing BCs \Hobc{L} proposed in
\cite{Buchman2006} are, for a flat-space background,
\begin{equation}
  \label{e:BL}
  {\mathcal B}_L:\quad
  [r^2 (\partial_t + \partial_{r})]^{L-1} (r^5 \Psi_0) \hateq 0,
\end{equation} 
where $\hateq$ denotes equality at the boundary. Here the Newman-Penrose
scalar $\Psi_0$ is evaluated for a null tetrad $(l^\alpha, k^\alpha,
m^\alpha, \bar m^\alpha)$ adapted to the background spacetime. The BCs
\eref{e:BL} are perfectly absorbing for all perturbations with angular
momentum numbers $\ell \leqslant L$.  We note that for $L=1$, \eref{e:BL}
reduces to the often-used freezing-$\Psi_0$ condition, $\partial_t
\Psi_0 \hateq 0$ \cite{Bardeen2002,Kidder2005,Sarbach2005,Rinne2007,
Scheel2006,Boyle2007,Scheel2008}. 
(We have not included the outer time derivative in \eref{e:BL};
the purpose of that time derivative is to address a static background
contribution to $\Psi_0$, but this contribution vanishes in flat space.)

Instead of $\Psi_0$, we choose to work with the RWZ scalars
$\Phi^{(\pm)}_{\ell m}$ describing gauge-invariant gravitational
perturbations about Schwarzschild spacetime (see \cite{Sarbach2001}
and references therein; of course, this includes our assumption of a 
flat-space background as a special case).  
The superscript $(\pm)$ refers to the parity
of the perturbations: $(+)$ for even and $(-)$ for odd parity.  The
indices $\ell m$ refer to a spherical harmonic decomposition and will
usually be suppressed in the following. We use the RWZ scalars because
they have the advantage that they obey a closed evolution equation, 
the RWZ equation, which in flat space reads
\begin{equation}
  \label{e:FlatRWZEq}
  \left[ \partial_t^2 - \partial_r^2 + \frac{\ell(\ell+1)}{r^2} \right]
  \Phi^{(\pm)} = 0.
\end{equation}
This equation arises from the scalar wave equation on $\Phi^{(\pm)}/r$
after a decomposition into spherical harmonics; it is also known as
the Euler-Poisson-Darboux equation \cite{Darboux1915}.

The relation between $\Phi^{(\pm)}$ and the perturbations 
$\delta \Psi_0$ of $\Psi_0$ is provided by equation 22 in 
\cite{Buchman2007}, which for a flat-space background reduces to
\begin{equation}
  \label{e:Psi0fromPhi} 
  \delta \Psi_0 = l^a l^b \tilde \nabla_a \tilde \nabla_b 
  [r (\Phi^{(+)} + i \Phi^{(-)})] 
    m^C m^D \hat \nabla_C \hat \nabla_D Y.
\end{equation}
Here $Y$ are the standard scalar spherical harmonics.
We have suppressed indices $\ell m$ on $Y$ and $\Phi^{(\pm)}$, which are being
summed over.  

Equation \eref{e:Psi0fromPhi} allows us to translate the BCs \eref{e:BL} on
$\Psi_0$ into BCs on $\Phi^{(\pm)}$,
\begin{equation}
  \label{e:FlatCL}
  [r^2 (\partial_t + \partial_r)]^{L+1} \Phi \hateq 0,
\end{equation}
which holds for both parities (and hence we suppress the superscript $(\pm)$).
We note that \eref{e:FlatCL} are the well-known Bayliss-Turkel
conditions \cite{Bayliss1980} for the scalar wave equation on
$\Phi/r$, equation \eref{e:FlatRWZEq}.


\subsection{The auxiliary system at the boundary}
\label{s:AuxVars}

The BCs \eref{e:FlatCL}, which are equivalent to the \Hobc{L} conditions
\eref{e:BL}, are difficult to implement numerically
because they contain higher-order radial derivatives. 
We follow a procedure developed for the scalar wave equation 
in \cite{Hagstrom1998,Huan2000,Givoli2001} in order to eliminate these
derivatives.
A set of auxiliary variables is introduced,
\begin{equation}
  \label{e:AuxVars}
  w_k \equiv r^{-(2k+1)}[r^2 (\partial_t + \partial_r)]^k \Phi,
\end{equation}
where again the parity $(\pm)$ and the indices $\ell m$ are suppressed.
Hence these auxiliary variables obey the recursion relation
\begin{equation}
    \label{e:AuxVarRecursion}
  \left(\partial_t + \partial_r + \frac{2k+1}{r}\right) w_k = w_{k+1}.
\end{equation}
Using the wave equation \eref{e:FlatRWZEq} and induction over $k$, we
can prove the identity \cite{Huan2000}
\begin{equation}
  \label{e:HuanIdentity}
  \left(\partial_t - \partial_r - \frac{1}{r}\right) w_k
  = \frac{1}{r^2} [-\ell(\ell+1) + k(k-1)] w_{k-1}.
\end{equation}
Equations \eref{e:AuxVarRecursion} and \eref{e:HuanIdentity} can be
combined to eliminate the radial derivatives,
\begin{equation}
  \label{e:AuxODEs}
  \left( \partial_t + \frac{k}{r} \right) w_k = 
  \frac{1}{2 r^2} [-\ell(\ell+1) + k(k-1)] w_{k-1} 
  + \half w_{k+1}.  
\end{equation}
The BC \eref{e:FlatCL} is equivalent to
\begin{equation}
  \label{e:wL+1BC}
  w_{L+1} \hateq 0,
\end{equation}
which closes the system of ODEs \eref{e:AuxODEs}.
We integrate \eref{e:AuxODEs} on the boundary for 
$1 \leqslant k \leqslant L$, substituting \eref{e:wL+1BC} and
$w_0 = \Phi/r$.


\subsection{Extraction of the RWZ scalars}
\label{s:Extraction}

The RWZ scalars need to be computed from the spacetime metric. 
We follow the gauge-invariant treatment of \cite{Sarbach2001},
restricted to a Minkowski background in standard coordinates. 
The starting point is a decomposition of the metric perturbations with
respect to scalar, vector and tensor spherical harmonics.
The basis harmonics are defined by
\begin{eqnarray}
  \label{e:YSharmonics}
  \fl Y_A \equiv \hat \nabla_A Y, \qquad S_A \equiv \hat \epsilon^B{}_A
  Y_B, \nonumber\\ \fl Y_{AB} \equiv [\hat \nabla_{(A} Y_{B)}]^\mathrm{TF} = 
  \hat \nabla_{(A} \hat \nabla_{B)} Y + \half \ell(\ell+1) \hat g_{AB} Y,
  \qquad S_{AB} \equiv \hat \nabla_{(A} S_{B)},
\end{eqnarray}
where $\mathrm{TF}$ denotes the tracefree part. 
The two parities are treated separately.

\subsubsection{Odd parity.}

Odd-parity perturbations of the spacetime metric are decomposed as
\begin{eqnarray}
  \label{e:OddDecomp1}
  \delta g_{Ab} &=& h_b S_A,\\
  \label{e:OddDecomp2}
  \delta g_{AB} &=& 2 k S_{AB}.
\end{eqnarray}
From the amplitudes $h_a$ and $k$, we construct the gauge-invariant
potential
\begin{equation}
  \label{e:hinv}
  h^{(\mathrm{inv})}_a = h_a - r^2 \tilde \nabla_a \left( \frac{k}{r^2} \right),
\end{equation}
i.e.
\begin{eqnarray}
  \label{e:hinvt}
  h^{(\mathrm{inv})}_t &=& h_t - \dot k,\\
  \label{e:hinvr}
  h^{(\mathrm{inv})}_r &=& h_r - r^2 \left( \frac{k}{r^2} \right)'.
\end{eqnarray}
Here and in the following, a dot (prime) denotes partial differentiation with
respect to $t$ ($r$).
The Regge-Wheeler scalar $\Phi^{(-)}$ is defined as
\begin{eqnarray}
  \label{e:FlatRWscalar}
  \fl \Phi^{(-)} = -\frac{r^3}{\lambda} \tilde \epsilon^{ab} \tilde \nabla_a
    \left(\frac{h^{(\mathrm{inv})}_b}{r^2}\right)
    = \frac{r^3}{\lambda} \left[ \partial_t \left( 
        \frac{h_r^{(\mathrm{inv})}}{r^2} \right) 
       - \partial_r \left( \frac{h_t^{(\mathrm{inv})}}{r^2} \right)
     \right] \nonumber\\
    = \frac{r}{\lambda} \left(\dot h_r - h_t' + \frac{2}{r} h_t \right),
\end{eqnarray}
where $\lambda \equiv (\ell-1)(\ell+2)$.
Equation \eref{e:FlatRWscalar} is valid for $\ell \geqslant 2$; 
the special case $\ell=1$
corresponds to a non-dynamical degree of freedom (variation of the
background angular momentum) that is not needed in our treatment. 

\subsubsection{Even parity.}

Even-parity perturbations are decomposed as 
\begin{eqnarray}
  \label{e:EvenDecomp1}
  \delta g_{ab} &=& H_{ab} Y, \\
  \delta g_{Ab} &=& Q_b Y_A, \\
  \label{e:EvenDecomp3}
  \delta g_{AB} &=& r^2 (K \hat g_{AB} Y + G Y_{AB}).
\end{eqnarray}
We define the gauge parameters
\begin{equation}
  p_a = Q_a - \half r^2 G_{|a},
\end{equation}
i.e.
\begin{eqnarray}
  p_t &=& Q_t - \half r^2 \dot G,\\
  p_r &=& Q_r - \half r^2 G'.
\end{eqnarray}
The Zerilli 1-form is given by 
\begin{equation}
  Z_a = H_{ab} r^{|b} - r K_{|a} - \half \ell(\ell+1) r G_{|a} +
  r^{|b} \omega_{ab} + 2 r v_{b|a} p^b
\end{equation}
with $\omega_{ab} \equiv 2 p_{[b|a]}$ and $v_a \equiv r^{|a}/r$, i.e.
\begin{eqnarray}
  \label{e:Zt}
  Z_t &=& H_{tr} - r \dot K - \half \ell(\ell+1) r \dot G + \dot p_r - p_t',\\
  \label{e:Zr}
  Z_r &=& H_{rr} - r K' - \half \ell(\ell+1) r G' - \frac{2}{r} p_r.
\end{eqnarray}
The gauge-invariant potential is
\begin{equation}
  K^{(\mathrm{inv})} = K + \half \ell(\ell+1) G - \frac{2}{r} r^{|b} p_b 
    = K + \half \ell(\ell+1) G - \frac{2}{r} p_r.
\end{equation}
Finally, we obtain the Zerilli scalar
\begin{equation}
\label{e:FlatZscalar}
  \fl \Phi^{(+)} = -\frac{r}{\lambda \ell(\ell+1)} (2 r^{|a} Z_a 
    + \lambda K^{(\mathrm{inv})})
    = -\frac{r}{\lambda \ell(\ell+1)} (2 Z_r + \lambda K^{(\mathrm{inv})}).
\end{equation}
Again, this formula is valid for $\ell \geqslant 2$; the non-dynamical
cases $\ell = 0$ (variation of the background mass) and $\ell = 1$ 
(pure gauge) are not needed.


\subsection{Boundary conditions for the generalized harmonic Einstein equations}
\label{s:EinsteinBCs}

Next we describe the BCs that we impose on the actual
Einstein equations that are being solved in the interior of the domain.
We consider a first-order formulation of these equations in generalized
harmonic coordinates (see \cite{Lindblom2006} and references therein). 
Generalized harmonic spacetime coordinates $x^\alpha$ are defined by
\begin{equation}
  \label{e:HarmonicGauge}
  \Box x^\alpha = -g^{\beta\gamma} \Gamma^\alpha{}_{\beta\gamma} = H^\alpha,
\end{equation}
where $\Box$ is the scalar d'Alembert operator of the spacetime metric 
$g_{\alpha\beta}$, $\Gamma^\alpha{}_{\beta\gamma}$ are its Christoffel symbols,
and $H^\alpha$ are freely specifiable gauge source functions.
The evolved variables are $g_{\alpha\beta}$ and their 
first derivatives
\begin{eqnarray} 
  \Pi_{\alpha\beta} &=& -t^\gamma \partial_\gamma g_{\alpha\beta},\\
  \Phi_{i\alpha\beta} &=& \partial_i g_{\alpha\beta},
\end{eqnarray}
where $t^\alpha$ is the future-directed unit timelike normal to the $t
= \mathrm{const}$ slices.
Greek indices $\alpha,\beta,\ldots$ are spacetime indices and Latin
indices $i,j,\ldots$ from the middle of the alphabet are spatial.
Boundary data must be specified for the incoming fields at the boundary,
\begin{equation}
  \label{e:IncomingFields}
  u^{1-}_{\alpha\beta} = \Pi_{\alpha\beta} - n^i \Phi_{i\alpha\beta} -
  \gamma_2 g_{\alpha\beta},
\end{equation}
where $n^\alpha$ is the outward-pointing unit spatial normal to the 
boundary on the $t = \mathrm{const}$ slices, $t^\alpha n_\alpha = 0$. 
The parameter $\gamma_2$ appears due to the addition of
constraint-damping terms to the evolution equations; let us also define
\begin{equation}
  \label{e:IncomingFieldsWithoutGamma2}
  \tilde u^{1-}_{\alpha\beta} \equiv u^{1-}_{\alpha\beta} + \gamma_2 g_{\alpha\beta}
  = \Pi_{\alpha\beta} - n^i \Phi_{i\alpha\beta}.
\end{equation}
Our BCs are of a similar type as those considered in 
\cite{Kreiss2006} but with boundary data derived from our
auxiliary system at the boundary as described in the following.

The incoming fields are split into three different pieces by
three projection operators $P^\mathrm{C}$, $P^\mathrm{P}$ and $P^\mathrm{G}$
(referring to constraint, physical and gauge BCs). 
In order to define them, we introduce the null vectors 
$l^\alpha \equiv (t^\alpha + n^\alpha)/\sqrt{2}$ and 
$k^\alpha \equiv (t^\alpha - n^\alpha)/\sqrt{2}$ and the spatial
metric induced on the boundary, 
$P_{\alpha\beta} \equiv g_{\alpha\beta} + t_\alpha t_\beta - n_\alpha n_\beta$.
The projection operators are given by
\begin{eqnarray}
  P^\mathrm{C}_\alpha{}^{\gamma\delta} &=& 2k^{(\gamma} \delta_\alpha{}^{\delta)} 
   - k_\alpha g^{\gamma\delta},\\
  P^\mathrm{P}_{\alpha\beta}{}^{\gamma\delta} &=& P_\alpha{}^\gamma
  P_\beta{}^\delta - \half P_{\alpha\beta} P^{\gamma\delta},\\
  P^\mathrm{G}_\alpha{}^{\gamma\delta} &=& \delta_\alpha{}^\gamma l^\delta.
\end{eqnarray}
The complements of the kernels of these three projection operators
are disjoint and together they span the space of all symmetric $2$-tensors.

We impose constraint-preserving BCs by requiring the generalized 
harmonic gauge constraint \eref{e:HarmonicGauge} to hold at the boundary. 
This can be written in the form
\begin{equation}
  \label{e:CBC}
  P^\mathrm{C}_\alpha{}^{\gamma\delta} \tilde u^{1-}_{\gamma\delta} 
  \hateq F^\mathrm{C}_\alpha,
\end{equation}
where $F^\mathrm{C}_\alpha$ is a function of outgoing and zero-speed
characteristic fields and gauge source functions, see equation 32 
of \cite{Rinne2007}.
The rank-2 projection operator $P^\mathrm{P}$ describes the BCs
 on the two gravitational degrees of freedom. 
They take the form
\begin{equation}
  \label{e:PBC}
  P^\mathrm{P}_{\alpha\beta}{}^{\gamma\delta} \tilde u^{1-}_{\gamma\delta}
  \hateq F^\mathrm{P}_{\alpha\beta}.
\end{equation}
Here $F^\mathrm{P}_{\alpha\beta}$ are boundary data that will be
constructed in the following subsection from our auxiliary variables 
at the boundary; thus these ``physical'' boundary data implement the 
absorbing BC \Hobc{L} that we want to impose.
The remaining BCs
\begin{equation}
  \label{e:GBC1}
  P^\mathrm{G}_\alpha{}^{\gamma\delta} \tilde u^{1-}_{\gamma\delta} 
  \hateq F^\mathrm{G}_\alpha
\end{equation}
are related to the residual gauge freedom within 
the generalized harmonic gauge:
we can still perform infinitesimal coordinate transformations of the
coordinates $x^\alpha \rightarrow x^\alpha + \xi^\alpha$ provided that
$\xi^\alpha$ obeys the wave equation.
The $F^\mathrm{G}_\alpha$ are free data that will be specified so that
our BCs are compatible with the exact solution that we
want to reproduce. 
In realistic simulations without an exact solution at hand, we would
take the $F^\mathrm{G}_\alpha$ to vanish. 

The BCs \eref{e:CBC}--\eref{e:GBC1} were proven in 
\cite{Kreiss2006,Kreiss2007} to yield a well-posed IBVP
for the Einstein equations in harmonic gauge.
This result cannot be directly applied to our formulation because
(i) whereas the second-order Einstein equations are considered in 
\cite{Kreiss2006,Kreiss2007}, our evolution system
\cite{Lindblom2006} is a first-order reduction thereof, and
(ii) rather than being {\it a priori} given functions of time,
the boundary data $F^\mathrm{P}_{\alpha\beta}$ in \eref{e:PBC} 
depend implicitly on the dynamical fields, as we shall see in 
the following subsection.
Nevertheless, it has been shown in \cite{Ruiz2007} that the
absorbing BCs \Hobc{L} are well posed for the second-order
Einstein equations in harmonic gauge at least in the 
high-frequency limit.
Therefore, it seems likely that our method leads to a well-posed IBVP; 
a proof is beyond the scope of this paper.


\subsection{Construction of absorbing boundary data}
\label{s:Reconstruction}

Finally, we show how to construct boundary data $F^\mathrm{P}_{\alpha\beta}$
in \eref{e:PBC} that correspond to \Hobc{L}. 
Recall that we are assuming that the fields can be linearized about
flat space close to the outer boundary, which is taken to be a
sphere $r = \mathrm{const}$.
Hence we have $t^\alpha = \delta_t{}^\alpha$ and 
$n^\alpha = \delta_r{}^\alpha$ for the background, and the 
incoming fields (without the $\gamma_2$ term) in 
\eref{e:IncomingFieldsWithoutGamma2} read
\begin{equation}
  \tilde u^{1-}_{\alpha\beta} = -(\partial_t + \partial_r) g_{\alpha\beta}. 
\end{equation}
The only non-vanishing components of the ``physical'' BCs
 \eref{e:PBC} are the angular components, $\alpha\beta = AB$.
We find
\begin{equation}
  \label{e:BdryDataGeneral}
  F^\mathrm{P}_{AB} \hateq P^\mathrm{P}_{AB}{}^{\gamma\delta} \tilde u^{1-}_{\gamma\delta} =
  -r^2(\partial_t + \partial_r) (r^{-2} \delta g_{AB}^\mathrm{TF}),
\end{equation}
where $\mathrm{TF}$ denotes the trace-free part with respect to the
metric $\hat g$ on the 2-sphere.
We now derive expressions for the right-hand side of the above equation
in the RWZ formalism that involve the auxiliary variables at the boundary.

\subsubsection{Odd parity.}

From \eref{e:hinvt} and \eref{e:hinvr} we obtain
\begin{equation}
  \label{e:ldothinv1}
  h_t^{(\mathrm{inv})} + h_r^{(\mathrm{inv})} = h_t + h_r 
  - r^2 (\partial_t + \partial_r) (r^{-2} k).
\end{equation}
On the other hand \cite{Sarbach2001}, the gauge-invariant potential 
is related to the Regge-Wheeler scalar via 
$h^{(\mathrm{inv})} = \tilde * \rmd(r \Phi^{(-)})$, where
$\tilde *$ denotes the Hodge dual with respect to $\tilde g$.
Thus,
\begin{equation}
  \label{e:ldothinv2}
  h_t^{(\mathrm{inv})} + h_r^{(\mathrm{inv})} = -(\partial_t +
  \partial_r)(r \Phi^{(-)}) = -r^2 w_1^{(-)} - r w_0^{(-)},
\end{equation}
where we have substituted the definition \eref{e:AuxVars} of the
auxiliary variables.
Combining \eref{e:ldothinv1} and \eref{e:ldothinv2},
\begin{equation}
  \label{e:OddBdryData1}
  r^2 (\partial_t + \partial_r) (r^{-2} k) = h_t + h_r + r^2 w_1^{(-)} + r w_0^{(-)}.
\end{equation}
Finally \eref{e:OddBdryData1} is substituted in the expression
\eref{e:BdryDataGeneral} for the boundary data 
(noting \eref{e:OddDecomp2}),
\begin{eqnarray}
  \label{e:OddBdryData2}
  \fl F^\mathrm{P}_{AB} &=&
  - r^2 (\partial_t + \partial_r)(r^{-2} \delta g_{AB}^\mathrm{TF}) 
  = -2 r^2 \left[ (\partial_t + \partial_r) (r^{-2} k) \right] S_{AB}\nonumber\\
  \fl &=& -2\left[ h_t + h_r + r^2 w_1^{(-)} + r w_0^{(-)}\right] S_{AB}.
\end{eqnarray}
Here we see clearly how the information from the auxiliary system
that implements \Hobc{L}
enters the boundary data for the Einstein equations.
We note that $S_{AB} = 0$ for $\ell \leqslant 1$,
which justifies our neglecting this special case.

\subsubsection{Even parity.}

From \eref{e:Zt} and \eref{e:Zr} we obtain
\begin{equation}
  \label{e:ldotZ1}
  Z_t + Z_r = -\tfrac{\lambda}{2} r (\partial_r + \partial_r) G 
  - r (\partial_t + \partial_r) K 
    + \dot Q_r - Q_t' - \tfrac{2}{r} Q_r + H_{tr} + H_{rr}.
\end{equation}
Using the definitions \cite{Sarbach2001} $d \zeta = Z$ and $\Phi^{(+)}
 = \zeta/\lambda$, we find the alternate expression
\begin{equation}
  \label{e:ldotZ2}
  Z_t + Z_r = \lambda (\partial_t + \partial_r) \Phi^{(+)}
  = \lambda r w_1^{(+)},
\end{equation}
where we have substituted the definition \eref{e:AuxVars} 
of the auxiliary variables.
Combining \eref{e:ldotZ1} and \eref{e:ldotZ2},
\begin{equation}
  \label{e:EvenBdryData1}
  \fl r^2 (\partial_t + \partial_r) G = -2 r^2 w_1^{(+)} 
  - \tfrac{2}{\lambda} r^2 (\partial_t + \partial_r) K +
  \tfrac{2r}{\lambda} (\dot Q_r - Q_t' - \tfrac{2}{r} Q_r + H_{tr} + H_{rr}).
\end{equation}
Finally \eref{e:EvenBdryData1} is substituted in the expression
\eref{e:BdryDataGeneral} for the boundary data 
(noting \eref{e:EvenDecomp3}),
\begin{eqnarray}
  \label{e:EvenBdryData2}
  \fl F^\mathrm{P}_{AB} &=&
  -r^2 (\partial_t + \partial_r) (r^{-2} \delta g_{AB}^\mathrm{TF}) = 
  -\left[ r^2 (\partial_t + \partial_r) G \right] Y_{AB}\nonumber\\
  \fl &=& -\left[-2 r^2 w_1^{(+)} 
    - \tfrac{2}{\lambda} r^2 (\partial_t + \partial_r) K +
    \tfrac{2r}{\lambda} (\dot Q_r - Q_t' - \tfrac{2}{r} Q_r + H_{tr} +
    H_{rr}) \right] Y_{AB}.
\end{eqnarray}
We note that $Y_{AB} = 0$ for $\ell \leqslant 1$,
which justifies our neglecting this special case.

\subsubsection{Application to Cauchy-perturbative matching.}

We remark that the auxiliary variables $w_k^{(\pm)}$ ($k=0,1$)
appearing on the right-hand sides of \eref{e:OddBdryData2} and
\eref{e:EvenBdryData2} could equally well be computed from a given
exterior solution $\Phi^{(\pm)}$ of the RWZ equations, using the definition
\eref{e:AuxVars} of the auxiliary variables.
Hence we have shown how to construct BCs for the generalized 
harmonic Einstein equations in the CPM approach.


\section{Numerical tests}

In this section, we present numerical tests of our formulation of
higher-order absorbing BCs \Hobc{L} derived in
section \ref{s:Formulation}.

We evolve initial data representing exact wavelike solutions of the {\em
  linearized\/} Einstein equations using a {\em fully nonlinear\/}
evolution code that implements \Hobc{L}.  We extract the
RWZ scalars at the boundary and compare them to the corresponding
exact linearized solutions in order to assess the amount of spurious
reflections from the boundary. Because we evolve the full nonlinear
equations but our initial data and the solutions with which we compare
are based on linearized theory, our comparisons will disagree 
because of nonlinear effects; we repeat each run at several different
amplitudes in order to separate these nonlinear effects from the 
boundary reflections we are studying.

In section \ref{s:MultipolarSoln}, we describe exact solutions of
linearized gravity representing multipolar 
(with angular momentum number $\ell$) gravitational waves in TT gauge.  
We compute the RWZ scalars
and demonstrate that \Hobc{L=\ell} is perfectly
absorbing for these solutions.  In section~\ref{s:InitialData}, we
describe our use of these solutions to set up initial data for
numerical evolution.  After briefly explaining our pseudo-spectral
numerical method (section \ref{s:NumEvMethod}) and our
treatment of gauge BCs (section \ref{s:BdyOnGauge}),
we discuss our numerical results in section \ref{s:MultipolarEvoln}.
Finally, in section \ref{s:ReflCoeff}, we estimate reflection
coefficients from our numerical evolutions with the various BCs and
compare them with the predictions of \cite{Buchman2006}.


\subsection{Multipolar gravitational wave solutions}
\label{s:MultipolarSoln}

Teukolsky \cite{Teukolsky1982} gave the explicit form of the metric
for a quadrupolar ($\ell = 2$) gravitational wave linearized about flat
space in the TT gauge.  
This solution has recently been generalized to arbitrary angular
momentum number $\ell$ in \cite{Rinne2008c}.
The solutions are characterized by freely specifiable mode functions 
$F_\mathrm{out}(r-t), \, F_\mathrm{in}(r+t)$ for even parity and 
$G_\mathrm{out}(r-t), \, G_\mathrm{in}(r+t)$ for odd parity.  
They describe outgoing and incoming waves, respectively. 
There is an independent solution for each $m$, 
$-\ell \leqslant m \leqslant \ell$.

We note that whereas the solution for the metric is real,
the spherical harmonics are complex for $m \neq 0$. 
As a consequence, the amplitudes appearing in the spherical harmonic 
decomposition, equations \eref{e:OddDecomp1}--\eref{e:OddDecomp2} and
\eref{e:EvenDecomp1}--\eref{e:EvenDecomp3}, as well as the RWZ
scalars $\Phi^{(\pm)}_{\ell m}$ obey the reality conditions
\begin{equation}
  \label{e:RealityCondition}
  u_{\ell m}^\star = (-1)^m u_{\ell,-m},
\end{equation}
where $\star$ denotes complex conjugation.
More precisely, let us consider the solution for a single mode 
$(\ell,m) = (\hat \ell, \hat m)$.
Then we may write for any of the aforementioned quantities
\begin{equation}
  u_{\ell m} = c_{\ell m} u(t,r), 
\end{equation}
where $u(t,r)$ is a real function and the $c_{\ell m}$ are complex 
coefficients given by
\begin{eqnarray}
  \textrm{if } \hat m \geqslant 0: \quad & c_{\hat \ell, \hat m} 
  = c_{\hat \ell, -\hat m} = \half, 
  \quad  & c_{\ell m} = 0 \textrm{ otherwise},\\
  \textrm{if } \hat m < 0: \quad & c_{\hat \ell, \hat m} = \tfrac{\rmi}{2}, \quad 
  c_{\hat \ell, -\hat m} = -\tfrac{\rmi}{2}, \quad & c_{\ell m} = 0 
  \textrm{ otherwise}.
\end{eqnarray}
In the following, we only display the real functions $u(t,r)$.
The $\ell=2$ solutions in \cite{Teukolsky1982,Rinne2008c} differ by an
overall constant factor; we use the normalization of \cite{Rinne2008c} 
throughout.

From the explicit form of the metric, equations 4 and 5 in
\cite{Rinne2008c}, we read off the amplitudes
\begin{eqnarray}
  H_{tt} = H_{tr} = 0, \qquad
  H_{rr} = A, \qquad 
  Q_t = 0, \nonumber \\
  Q_r = r B, \qquad
  K = -\half A, \qquad
  G = C
\end{eqnarray}
for even parity and
\begin{equation}
  h_t = 0, \qquad 
  h_r = r K, \qquad
  k = \half r^2 L
\end{equation}
for odd parity, where the quantities $A,B,C$ and $K,L$ on the right-hand
sides are the radial functions defined in equations 8 and 9 in
\cite{Rinne2008c}.

Next we compute the RWZ scalars as described in section \ref{s:Extraction}. 
For brevity, we focus on $\ell=2$.
In terms of the mode functions, the outgoing solution is found to be
\begin{eqnarray}
  \label{e:RWZAnalyticTeukolsky1}
  \Phi^{(+)} &=& F_\mathrm{out}^{(4)} 
    - \frac{3 F_\mathrm{out}^{(3)}}{r} + \frac{3 F_\mathrm{out}^{(2)}}{r^2} ,\\
  \label{e:RWZAnalyticTeukolsky2}
  \Phi^{(-)} &=& G_\mathrm{out}^{(3)} 
    - \frac{3 G_\mathrm{out}^{(2)}}{r} + \frac{3 G_\mathrm{out}^{(1)}}{r^2},
\end{eqnarray}
where $F^{(k)}(x) \equiv d^k F(x) / dx^k$.
It can be verified easily that the RWZ scalars given in 
\eref{e:RWZAnalyticTeukolsky1} and \eref{e:RWZAnalyticTeukolsky2}
obey the RWZ equation \eref{e:FlatRWZEq}.

The auxiliary variables \eref{e:AuxVars} for the outgoing solution 
are found to be
\begin{equation}
  w_1^{(+)} = \frac{3F^{(3)}_\mathrm{out}}{r^3} 
    - \frac{6 F^{(2)}_\mathrm{out}}{r^4}, \quad
  w_2^{(+)} = \frac{6 F^{(2)}_\mathrm{out}}{r^5}, \quad
  w_3^{(+)} = 0,
\end{equation}
and similarly for odd parity.  
We see that the BC \Hobc{2}
corresponding to $w_3^{(+)} \hateq 0$ (equation \eref{e:wL+1BC})
is perfectly absorbing for this solution
(it is satisfied identically by the outgoing solution), whereas the
freezing-$\Psi_0$ condition \Hobc{1} 
corresponding to $w_2^{(+)} \hateq 0$
is not, although the correction in $w_2^{(+)}$ falls off quite fast 
with boundary radius ($\sim r^{-5}$).  
More generally for arbitrary $\ell$, \Hobc{L=\ell}
is perfectly absorbing whereas \Hobc{L<\ell} is not.


\subsection{Initial data for numerical evolutions}
\label{s:InitialData}

The initial data for our numerical evolutions
are taken from the exact linearized multipolar wave
solutions described in section \ref{s:MultipolarSoln}, evaluated at $t=0$.
We shall consider the cases $\ell=2,3,4$. 
For definiteness, we choose $m=2$ throughout; we have checked that our results
are insensitive to the particular choice of $m$.

\begin{figure}
\centering
\includegraphics[width=0.5\textwidth]{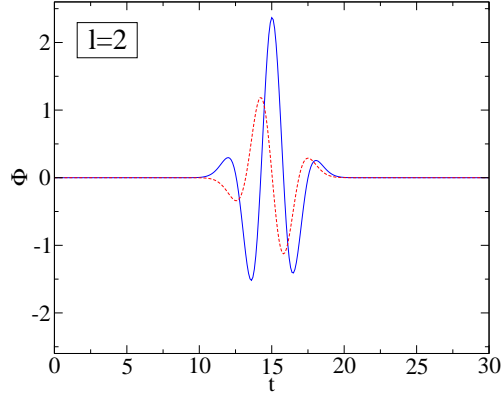}
\caption{\label{f:RWZanalytic} 
  Outgoing linearized gravitational wave with multipolar indices $\ell=m=2$ 
  and mode function \eref{e:GaussianModeFct} with parameters 
  $A=1$, $r_0 = 15$ and $\sigma = 1.5$. 
  Shown are the exact solutions \eref{e:RWZAnalyticTeukolsky1}
  for the Zerilli scalar $\Phi^{(+)}$ (solid) and
  \eref{e:RWZAnalyticTeukolsky2} for the Regge-Wheeler scalar 
  $\Phi^{(-)}$ (dashed) as functions of time, evaluated at $R = 30$.}
\end{figure}

We take an outgoing wave with mode function
\begin{equation}
  \label{e:GaussianModeFct}
  F_\mathrm{out}(x) = A \exp \left[ -\frac{(x-r_0)^2}{\sigma^2} \right] .
\end{equation}
The parameters are taken to be $r_0 = 15$ and $\sigma = 1.5$.  These
choices are dictated by the requirement that at the outer boundary
radius $R$ of our computational domain, which we place at $R=30$, the
wave initially vanish to numerical roundoff error; this guarantees
that the initial data are consistent with the boundary data at $R=30$.
Figure \ref{f:RWZanalytic} shows the exact solutions for the RWZ
scalars (for $\ell=2$) as functions of time evaluated at the outer
boundary $r=R=30$.

Although we will evolve the full nonlinear Einstein equations, we do
not solve the full nonlinear constraints initially (of course the
exact solution satisfies the constraints to linear order).  This is
because we are interested only in comparing the numerical solution
with the exact solution to linear order in perturbation theory.


\subsection{Numerical evolution method}
\label{s:NumEvMethod}

Our numerical implementation uses the Caltech-Cornell Spectral
Einstein Code (SpEC), which is based on a pseudo-spectral collocation
method described in more detail for example in \cite{Kidder2005}.
We evolve the first-order generalized harmonic form of the Einstein
equations as described in~\cite{Lindblom2006}.
The gauge-source functions $H^\alpha$ in \eref{e:HarmonicGauge} are
taken to be zero when that equation is evaluated in Cartesian components, 
compatible with the TT gauge of the exact linearized solution.

The computational domain for the evolutions presented here is a sphere
of radius $R=30$. 
It is subdivided into a small central sphere of radius $\Delta r = 7.5$ 
surrounded by three spherical shells, each of extent $\Delta r$.
In the spherical shells, 
each Cartesian tensor component of each tensor field is expanded in
Chebyshev polynomials in the radial direction and in spherical
harmonics $Y_{\ell m}(\theta,\phi)$ in the angular directions.
In the central sphere, the numerical solution is expanded in
the basis functions introduced by Matsushima and 
Marcus~\cite{Matsushima-Marcus:1995} with parameters $\alpha=1$ and $\beta=2$:
each Cartesian tensor component is expanded as $f(r,\theta,\phi)=\sum_{n\ell m} 
f_{n\ell m} Q_{n\ell}(r) Y_{\ell m}(\theta,\phi)$, 
where $Q_{n\ell}(r)$ depends on the index
$\ell$ and $Q_{n\ell}(r)\sim r^\ell$ near the origin.
The number of radial expansion coefficients is the same in each subdomain,
and is denoted by $N_r$. Likewise, the highest retained spherical harmonic
index is the same in each subdomain, and is denoted by $N_\ell$.
In all subdomains we retain all spherical harmonic $m$ coefficients
with $|m|\leq \ell$.

For the RWZ formalism, we need to compute the expansion coefficients 
of various tensors with respect to the basis $Y, Y_A, S_A, Y_{AB}, S_{AB}$
defined in \eref{e:YSharmonics}.
The SpEC code is already capable of computing expansions in
\emph{spin-weighted} spherical harmonics.
The conversion between the two bases is detailed in \ref{s:SpinWeighted}.
We note that there is no need to use spin-weighted harmonics, it is 
simply convenient in the SpEC code.

The evolution equations are integrated in time using a fourth-order 
Runge-Kutta scheme, with a Courant factor $\Delta t / \Delta x_\mathrm{min}$ 
of at most $2.25$, where $\Delta x_\mathrm{min}$ is the smallest distance
between two neighbouring collocation points. (This method of
defining a Courant factor for a pseudospectral code with uneven grid spacing
is only a crude approximation; so in practice, numerical stability
can be achieved with Courant factors larger than unity.)
As described in~\cite{Kidder2005}, the top four coefficients
in the \emph{tensor} spherical harmonic
expansion of each of our evolved quantities are set to
zero after each time step; this eliminates an instability associated with
the inconsistent mixing of tensor spherical harmonics in our approach.
The auxiliary ODEs \eref{e:AuxODEs} are integrated using the same
Runge-Kutta scheme.

The BCs on the incoming fields of the generalized harmonic
Einstein equations at the outer boundary and at the internal boundaries 
between neighbouring subdomains are enforced using a penalty 
method~\cite{Hesthaven2000}. At the internal boundaries, the
incoming fields on each boundary are set (weakly via the penalty method)
equal to the corresponding outgoing field on the neighbouring boundary.
At the outer boundary, we always impose the constraint-preserving boundary
conditions~(\ref{e:CBC}) on the constrained degrees of freedom, and we
vary the BCs on the physical and gauge degrees of freedom
as described below.


\subsection{Boundary conditions on gauge fields}
\label{s:BdyOnGauge}

Besides the BCs on the physical degrees of freedom
that are the subject of this paper, and those on the constraint
degrees of freedom for which we choose equation \eref{e:CBC}, 
we must also specify a BC \eref{e:GBC1} on the gauge degrees of freedom. 
The multipolar gravitational waves described in
section~\ref{s:MultipolarSoln} are in TT gauge,
\begin{equation}
  \delta g_{tt} = 0, \qquad \delta g_i{}^i = 0, \qquad
  \mathring{\nabla}_j \delta g^{ij} = 0,
\end{equation}
where $\mathring{\nabla}$ is the flat-space covariant derivative.
This implies that the harmonic gauge condition 
\eref{e:HarmonicGauge} with vanishing gauge source functions $H^\alpha$
---when evaluated in Cartesian components--- 
is satisfied in linearized theory.
However, the gauge boundary data $F^\mathrm{G}_\alpha$ in \eref{e:GBC1}
do not vanish for this solution; we have to specify them explicitly.
One might be tempted to replace the gauge BC \eref{e:GBC1} with
\begin{equation}
  t^\alpha u^{1-}_{\alpha\beta} = 0,
\end{equation}
which is compatible with TT gauge. 
Unfortunately, the ensuing IBVP is ill posed \cite{Rinne2006}.
Alternatively, one might hope to transform the solution to a
different gauge such that both the harmonic gauge condition and the
homogeneous version of the gauge BC \eref{e:GBC1}
are satisfied; however, it turns out that this cannot be achieved
for a purely outgoing solution \cite{RinneSarbachPrivComm2007}.

Hence we have decided to specify gauge boundary data $F^\mathrm{G}_\alpha$
that are computed from the exact linearized solution.
Alternatively, we set $F^\mathrm{G}_\alpha = 0$.
The latter is compatible with the exact solution at early times provided 
the initial data have compact support and the outer boundary is sufficiently 
far out. 
Once the outgoing wave hits the outer boundary, there will be a (small)
reflection which is pure gauge and hence should not affect the physical 
gravitational radiation as represented by the RWZ scalars.


\subsection{Numerical results}
\label{s:MultipolarEvoln}

Here we describe the results of evolving initial data describing exact
solutions of the {\em linearized} Einstein equations 
(section \eref{s:MultipolarSoln}) using our {\em fully nonlinear} 
evolution code that includes our new absorbing BCs.
Because we will compare the RWZ scalars extracted from our nonlinear
numerical evolution with their exact (linearized) values, our
comparisons will disagree because of nonlinear effects. Therefore we
repeat each run at several different amplitudes in order to separate
these nonlinear effects from the boundary reflections of interest.
The numerical resolution $(N_r, N_\ell)$ is chosen high enough such
that the numerical truncation error is
negligible compared to the differences we are studying.

We begin with the quadrupolar solution ($\ell=2$).
Here we consider amplitudes in the range $10^{-3.49} \leqslant A
\leqslant 10^{-1.39}$.
We performed a convergence test which showed that for any amplitude 
in this range, a numerical resolution of $N_r = 81$, $N_\ell=8$ 
is sufficient to remove the effects of numerical truncation error.

\begin{figure}
\includegraphics[width=0.5\textwidth]{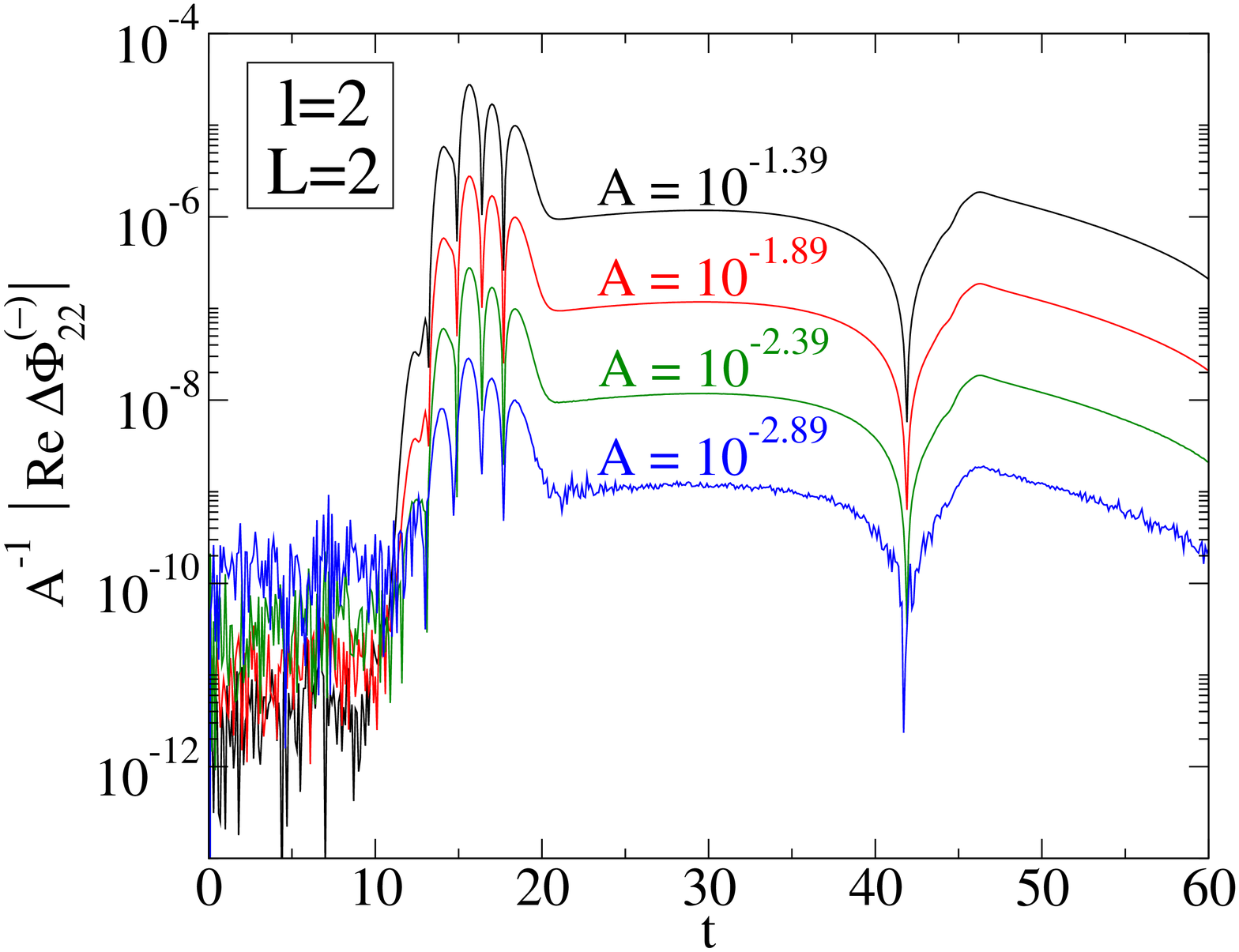} 
\includegraphics[width=0.5\textwidth]{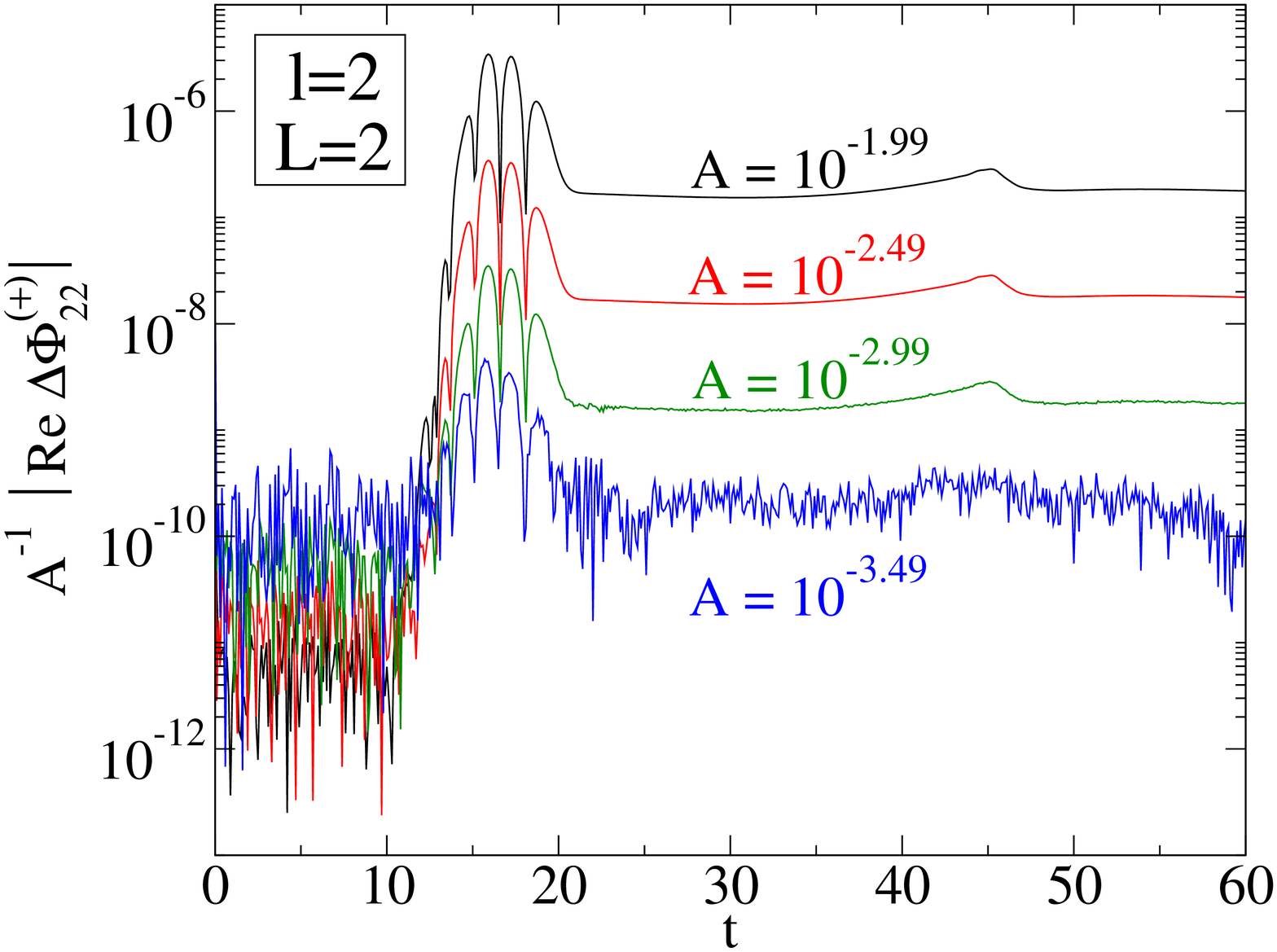}
\caption{\label{f:L2DeltaPhi22K2Analytic} 
  Difference between the numerically-extracted
  RWZ scalar $\Phi_{22}$ and its exact linearized value, evaluated on 
  the outer boundary, and divided by the amplitude $A$.  
  The exact linearized solution and the initial data for the evolution 
  contain only the $\ell=2, m=2$ mode of odd (left panel) or 
  even (right panel) parity.  
  The evolution uses \Hobc{2} and analytic gauge BCs. 
  Roundoff effects become visible at the smallest amplitude.
}
\end{figure}

First we consider \Hobc{2}, which was shown in 
section \ref{s:MultipolarSoln} to be perfectly absorbing for 
the $\ell=2$ solution.  
As mentioned in section \ref{s:BdyOnGauge}, there is some
freedom in the choice of gauge boundary data $F^\mathrm{G}_\alpha$.
First we compute these functions $F^\mathrm{G}_\alpha$ from the exact
solution; with this choice of boundary gauge, the entire spacetime
metric should agree with the exact linearized solution.  We compute
from the numerical evolution the $\ell=m=2$ component of the RWZ scalars,
evaluated at the outer boundary, for both an odd-parity and an
even-parity wave (this is the only nonvanishing component in the exact
solution, up to the reality condition \eref{e:RealityCondition}).  
We then compute the same quantity from the exact linearized solution, and
plot the difference in figure \ref{f:L2DeltaPhi22K2Analytic}.  
This difference has been normalized by the amplitude $A$, so that if the
quantity plotted decreases at least linearly with decreasing amplitude, 
the difference between the numerical and exact solution decreases
at least quadratically with amplitude, i.e.~the two solutions agree
to linear order in perturbation theory.
This is indeed what we see in figure \ref{f:L2DeltaPhi22K2Analytic}
(the decay exponent of the normalized difference is found to be 
close to $2$).
Similar behaviour is found for the remaining components of the RWZ 
scalars, which vanish for the exact solution.

\begin{figure}
\includegraphics[width=0.5\textwidth]{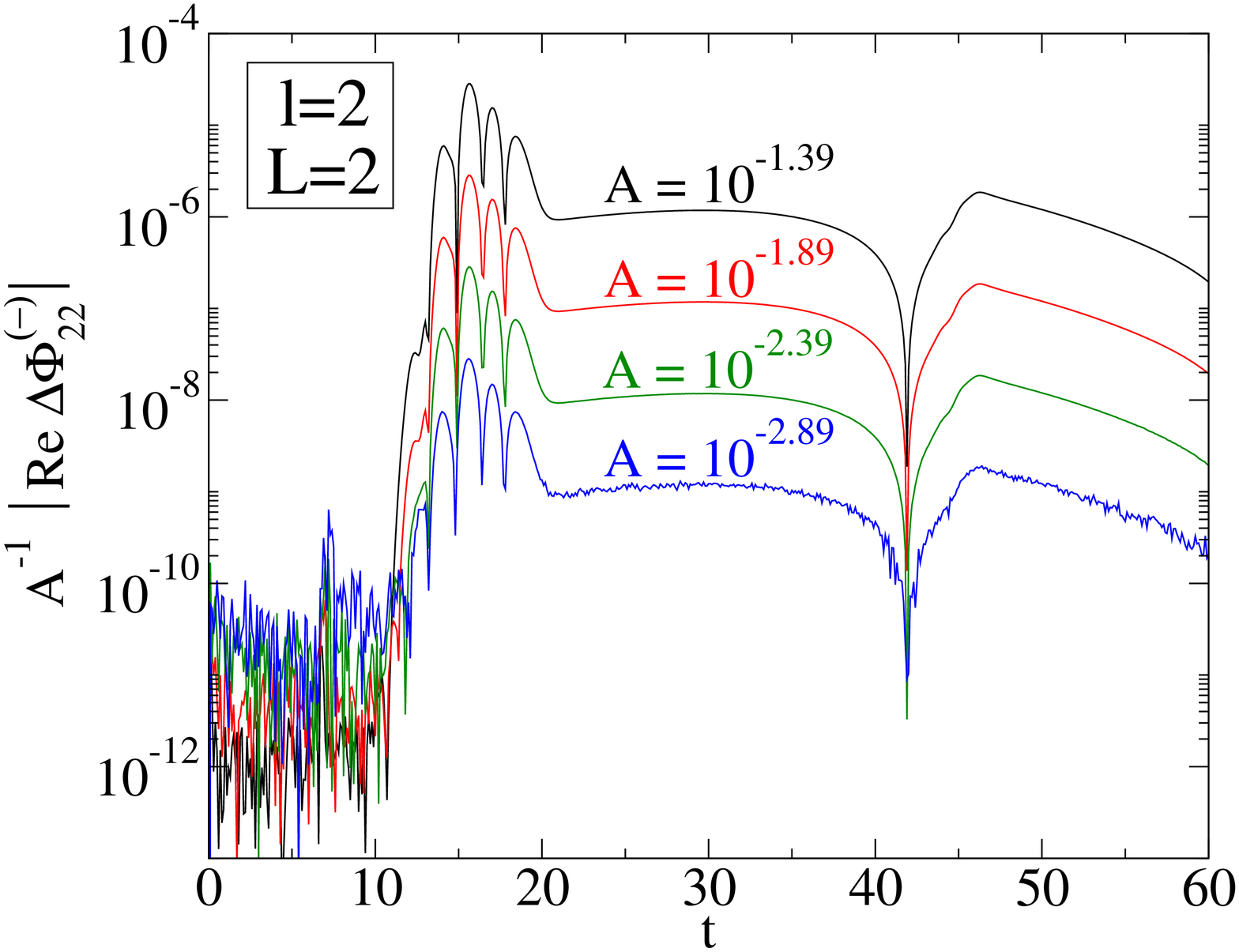} 
\includegraphics[width=0.5\textwidth]{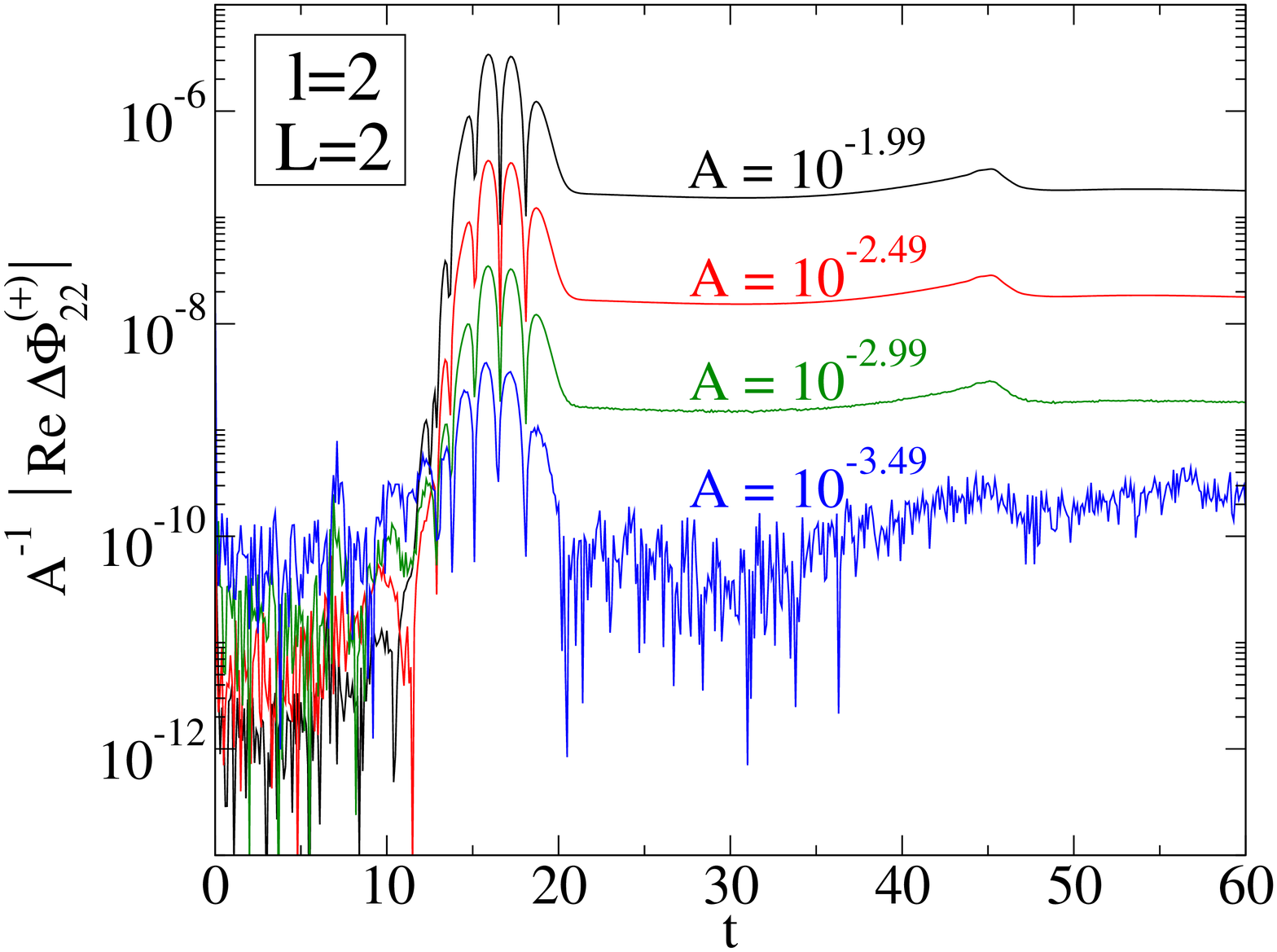}
\caption{\label{f:L2DeltaPhi22K2Freezing} 
  Same as figure~\ref{f:L2DeltaPhi22K2Analytic}, except freezing gauge BCs
  are used in the evolution.
}
\end{figure}

Next we replace the analytic gauge BC with the homogeneous one,
$F^\mathrm{G}_\alpha = 0$, also referred to as a \emph{freezing}
gauge BC.  The results are virtually unchanged (figure
\ref{f:L2DeltaPhi22K2Freezing}).  This demonstrates that the ambiguity
in the choice of gauge BC has no effect on the gauge-invariant
quantities. In all further evolutions discussed below, we
will continue to use the freezing gauge BC.

\begin{figure}
\includegraphics[width=0.5\textwidth]{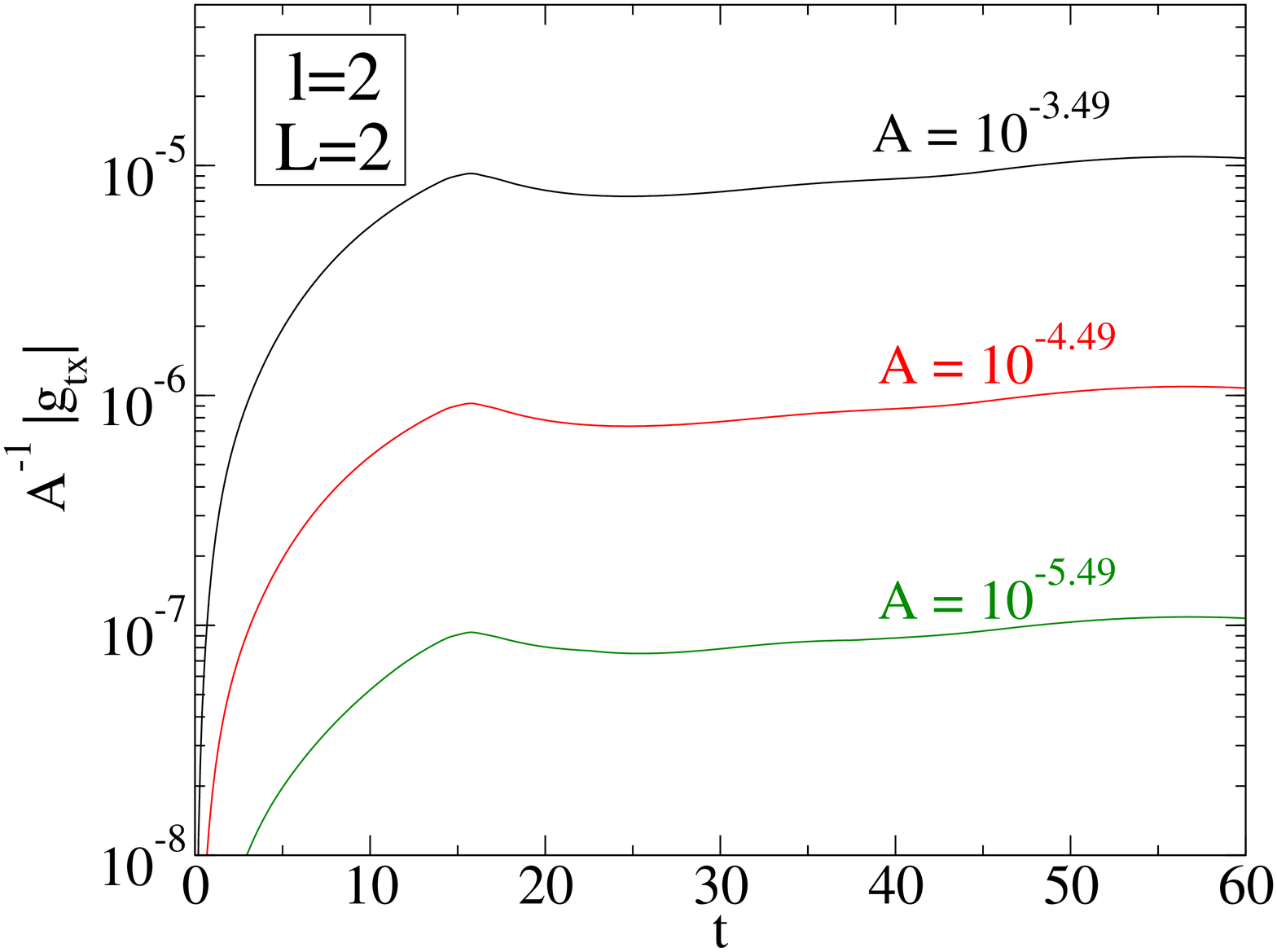} 
\includegraphics[width=0.5\textwidth]{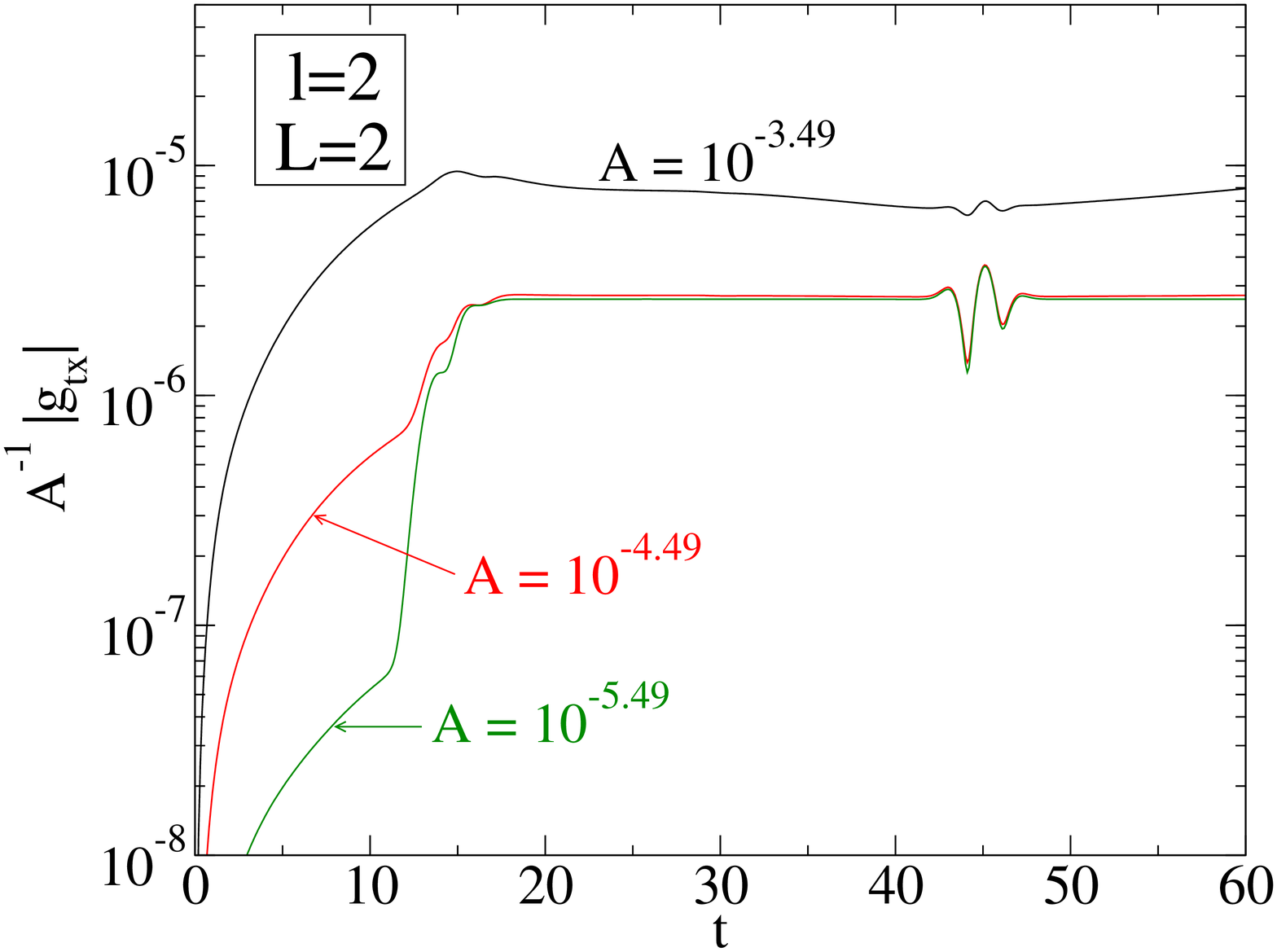}
\caption{\label{f:L2DeltapsitxK2}
  The component $g_{tx}$ of the metric, normalized by the amplitude $A$,
  for the even-parity evolutions shown in
  figures \ref{f:L2DeltaPhi22K2Analytic} and~\ref{f:L2DeltaPhi22K2Freezing}.
  The two panels are identical except for the gauge BCs used: analytic
  (left panel) and freezing (right panel).
}
\end{figure}

As discussed in section \ref{s:MultipolarSoln}, we expect the
freezing gauge BC to cause a small gauge wave reflection.
This should show up in gauge-dependent quantities such as the
spacetime metric.
Figure \ref{f:L2DeltapsitxK2} shows the component $g_{tx}$ 
of the metric, which vanishes for the exact solution because of the 
TT gauge.
Indeed this quantity is found to decay quadratically with amplitude when
analytical gauge BCs are used (left panel of 
figure \ref{f:L2DeltapsitxK2}; note again the quantity plotted
has been normalized by the amplitude).
For the freezing gauge BC (right panel), the curves begin to
overlap for sufficiently small amplitudes.
This indicates that the numerical solution differs from the exact 
solution to leading (linear) order in perturbation theory.
We had to use somewhat smaller amplitudes for this plot
($10^{-5.49} \leqslant A \leqslant 10^{-3.49}$) 
so that the nonlinear effects are smaller than the gauge wave
reflection that we are looking for.

\begin{figure}
\includegraphics[width=0.5\textwidth]{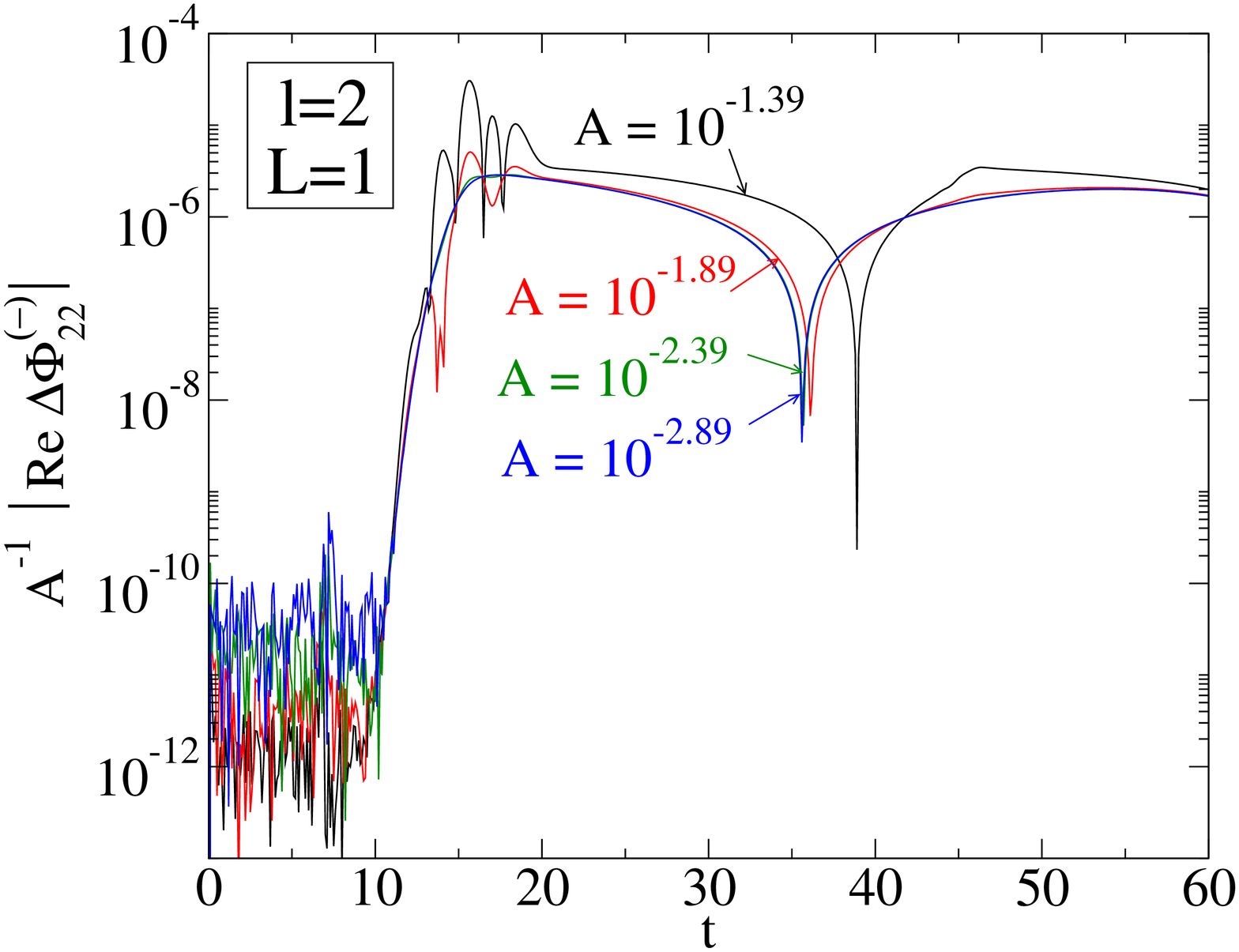} 
\includegraphics[width=0.5\textwidth]{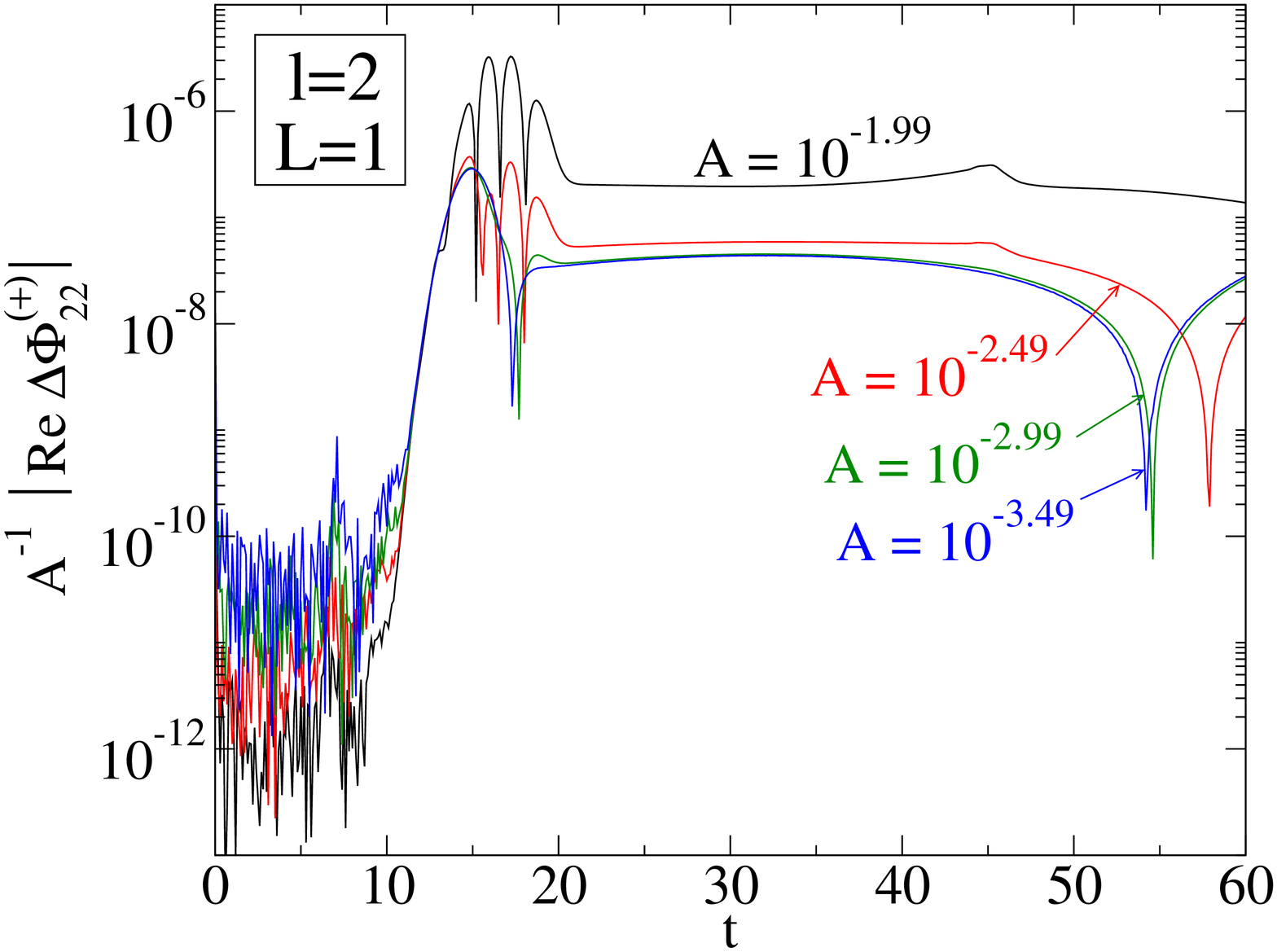}
\caption{\label{f:L2DeltaPhi22K1Freezing} 
  Same as figure \ref{f:L2DeltaPhi22K2Freezing}, except the evolution
  uses BC \Hobc{1}.
}
\end{figure}

We now repeat the evolution shown in figure \ref{f:L2DeltaPhi22K2Freezing}, 
but using \Hobc{1}; this corresponds to 
freezing $\Psi_0$ at the boundary. 
As shown analytically in section \ref{s:MultipolarEvoln}, \Hobc{1}
is \emph{not} compatible with the exact outgoing solution.
Figure \ref{f:L2DeltaPhi22K1Freezing} shows the RWZ scalars for
this numerical evolution.
The curves corresponding to smaller amplitudes nearly overlap, which implies
that the numerical solution differs from the exact one at linear order.
Note however that the normalized difference
is still relatively small (about $10^{-6}$ or $10^{-7}$, depending on 
the parity).
The remaining modes of the RWZ scalars not shown here behave in the
same way as for \Hobc{2}, i.e.~they decay at 
least quadratically with amplitude (note these modes vanish for the
exact linearized solution).

\begin{figure}
\includegraphics[width=0.5\textwidth]{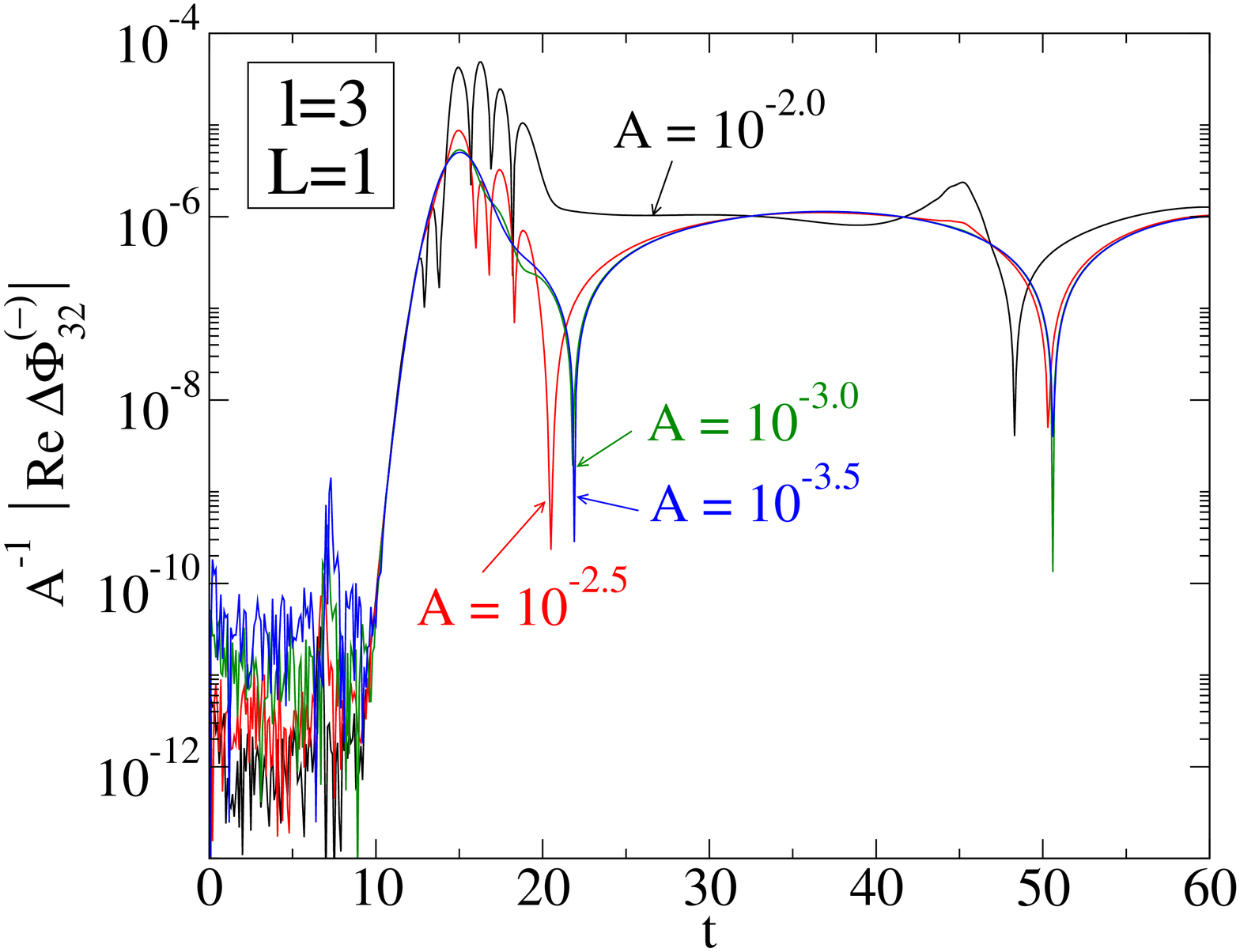} 
\includegraphics[width=0.5\textwidth]{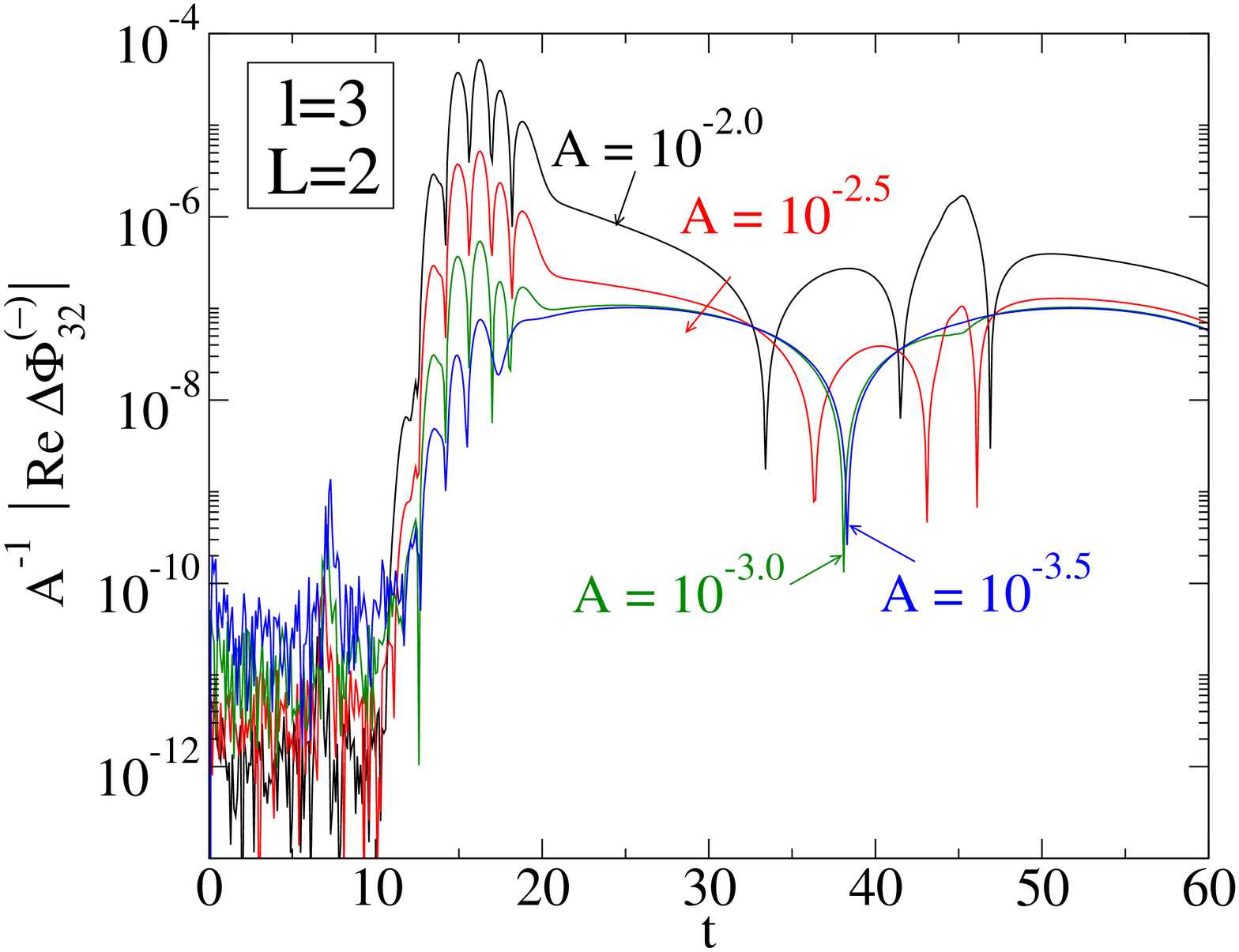}
\centerline{\includegraphics[width=0.5\textwidth]{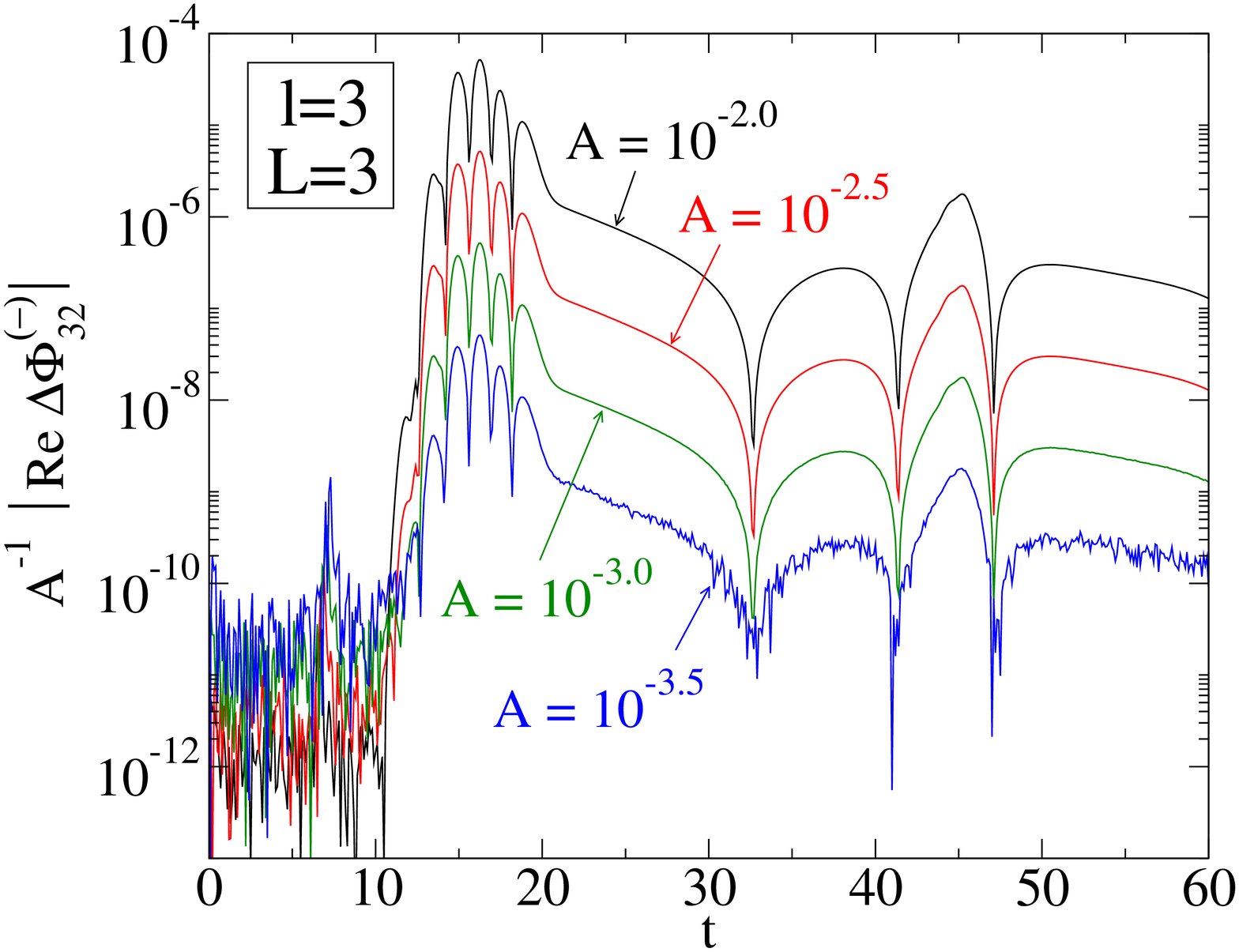}}
\caption{\label{f:L3DeltaPhi} 
  Difference between the numerically-extracted
  RWZ scalar $\Phi^{(-)}_{32}$ and its exact linearized value, 
  evaluated on the outer boundary, and divided by the amplitude $A$.  
  The exact linearized solution
  and the initial data for the evolution contain only the $\ell=3, m=2$ 
  odd-parity mode. The different panels correspond to absorbing BCs
  \Hobc{L} with $L=1$, $2$, and $3$. Freezing gauge BCs are used.
}
\end{figure}

\begin{figure}
\centerline{\includegraphics[width=0.5\textwidth]{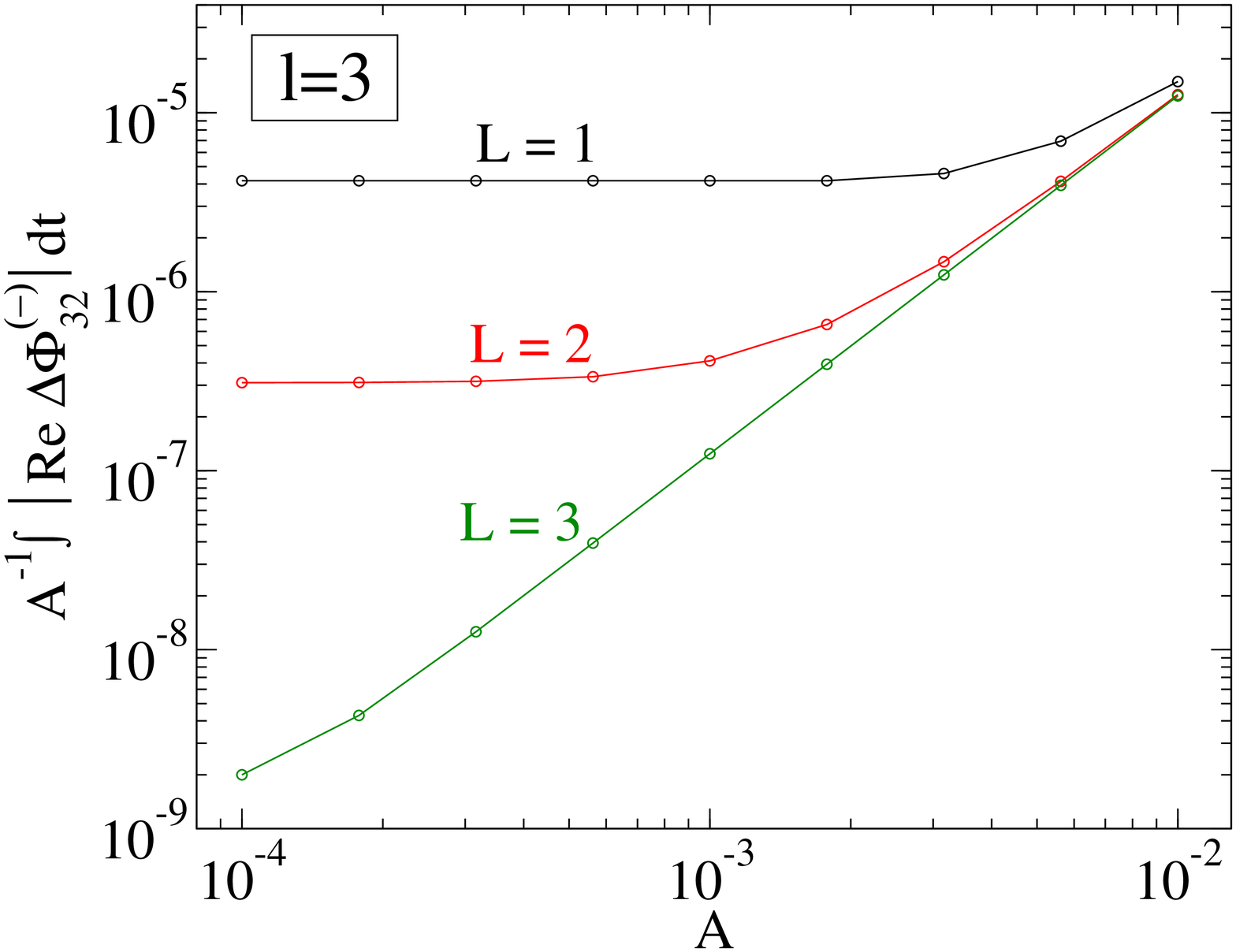}} 
\caption{\label{f:L3IntegralErrVsAmp} 
  Time integral of the curves shown in figure \ref{f:L3DeltaPhi},
  plotted against the amplitude $A$.
  The different curves correspond to absorbing BCs \Hobc{L} with
  $L=1$, $2$ and $3$. 
}
\end{figure}

We now turn to evolutions of octupolar ($\ell=3$) waves.
Here we consider amplitudes in the range $10^{-3.5} \leqslant A
\leqslant 10^{-2}$, and we find that a numerical resolution of $N_r
= 91$, $N_\ell = 10$ is sufficient to remove the effects of numerical 
truncation error for the amplitudes in this range.
Figure \ref{f:L3DeltaPhi} shows evolutions using
\Hobc{1}, \Hobc{2}, and \Hobc{3}. (We continue to use freezing gauge BCs.)
As expected from the analytic discussion in section \ref{s:MultipolarSoln},
\Hobc{3} is found to be consistent with the exact
linearized solution, whereas \Hobc{1} and \Hobc{2} are not.
To show the dependence on the order of the BC more clearly,
in figure \ref{f:L3IntegralErrVsAmp} we plot the time integrals of the 
curves in figure \ref{f:L3DeltaPhi} versus the amplitude.
We see clearly how for \Hobc{1} and \Hobc{2},
this time integral saturates in the limit of small amplitudes, 
at a level which is lower for \Hobc{2} than for \Hobc{1}.  
This saturation to a horizontal line at small amplitudes 
again indicates a difference between the numerical and exact
solution that decays only linearly with amplitude.
The reason why these curves increase for larger amplitudes is because nonlinear
effects become important.
The curve for \Hobc{3} behaves differently: it continues to decay with
decreasing amplitude. (The slight levelling off of this curve at the
smallest amplitude is caused by numerical round-off effects.)  
This is consistent with the theoretical prediction that
\Hobc{3} is perfectly absorbing for waves with $\ell \leqslant 3$, 
so the difference between the numerical and exact linear solution is (almost)
entirely due to nonlinear terms.  
Again, we find a nearly quadratic decay of the \Hobc{3} curve in 
figure \ref{f:L3IntegralErrVsAmp}, corresponding to a nearly cubic decay of
the difference between the numerical and exact solution.

\begin{figure}
\includegraphics[width=0.5\textwidth]{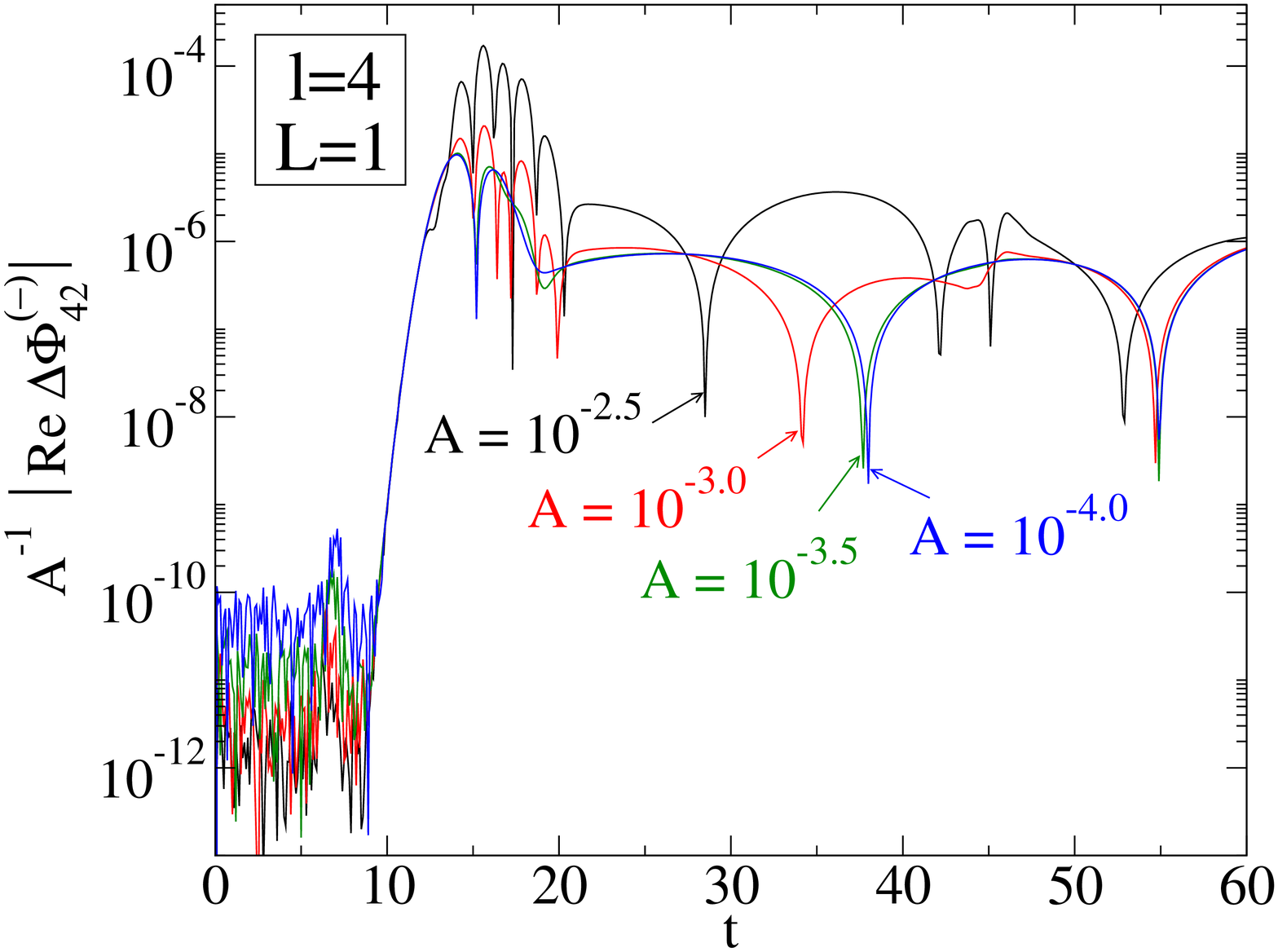}
\includegraphics[width=0.5\textwidth]{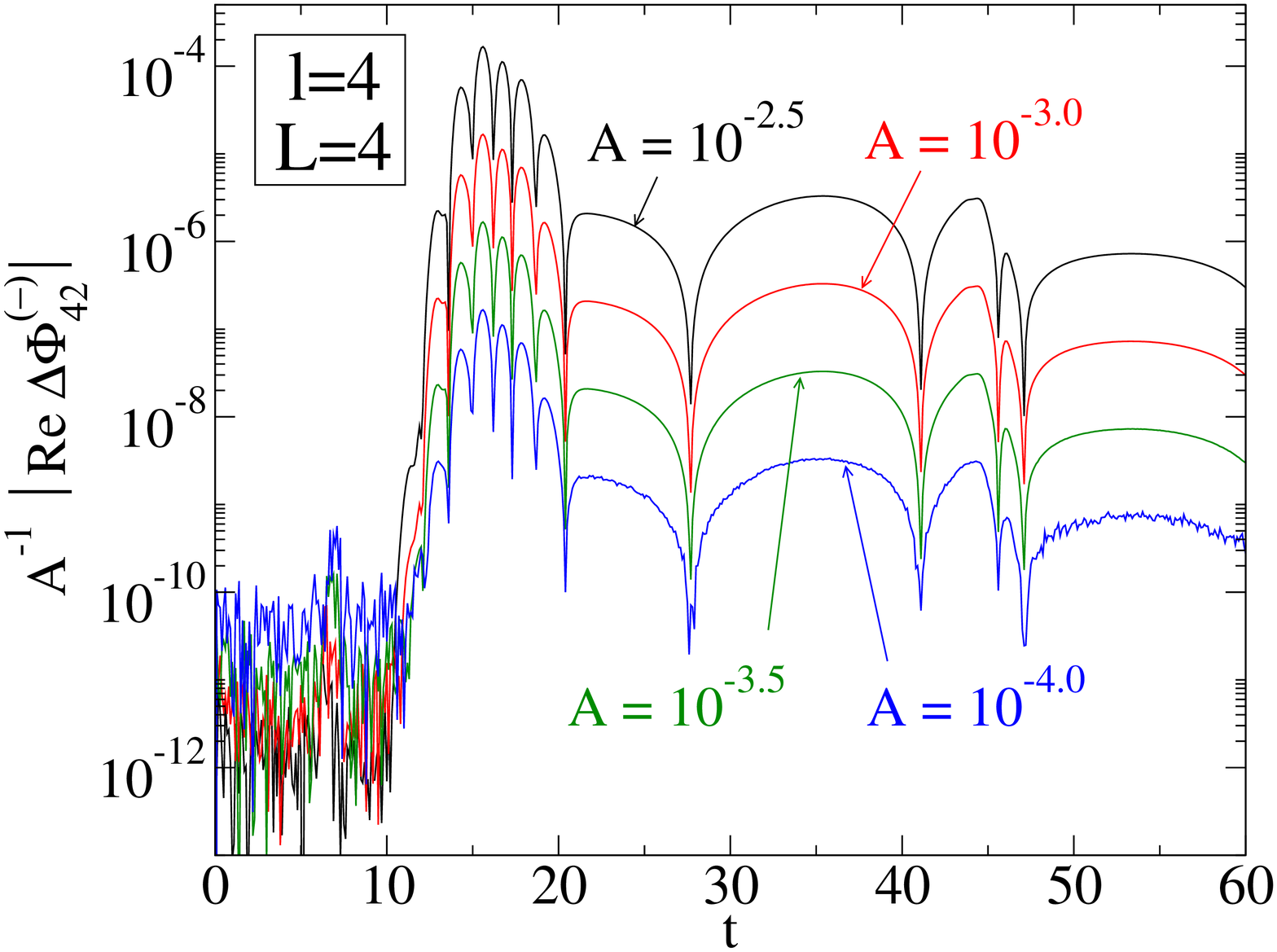}
\caption{\label{f:L4DeltaPhi} 
  Same as figure~\ref{f:L3DeltaPhi}, except the 
  exact linearized solution and the initial data for the evolution 
  now contain only the odd-parity $\ell=4, m=2$ mode.
  The different panels correspond to \Hobc{L} with $L=1$ and $4$.
}
\end{figure}

Finally we evolve the $\ell=4$ solution.
Here we consider amplitudes in the range $10^{-4} \leqslant A
\leqslant 10^{-2.5}$ and the numerical resolution is 
$N_r = 101$, $N_\ell = 12$.
Figure \ref{f:L4DeltaPhi} compares \Hobc{1}
(corresponding to freezing $\Psi_0$ at the boundary) and \Hobc{4} (which
is predicted to be perfectly absorbing for this solution).
Again the numerical results confirm that \Hobc{4} is compatible with the
exact linearized solution whereas \Hobc{1} is not.

\pagebreak
\subsection{Comparison with the predicted reflection coefficients}
\label{s:ReflCoeff}

In \cite{Buchman2006}, solutions to the linearized Bianchi equations
with higher-order absorbing BCs equivalent to \Hobc{L} 
were studied theoretically.  For monochromatic radiation,
analytical expressions for the reflection coefficients relating the
amplitude of an outgoing wave to the spurious reflected ingoing wave
were derived (equation 96 of \cite{Buchman2006}).  In this section, we
compare these predicted reflection coefficients with the numerical
results presented in section \ref{s:MultipolarEvoln}.

\begin{figure}
\includegraphics[width=0.5\textwidth]{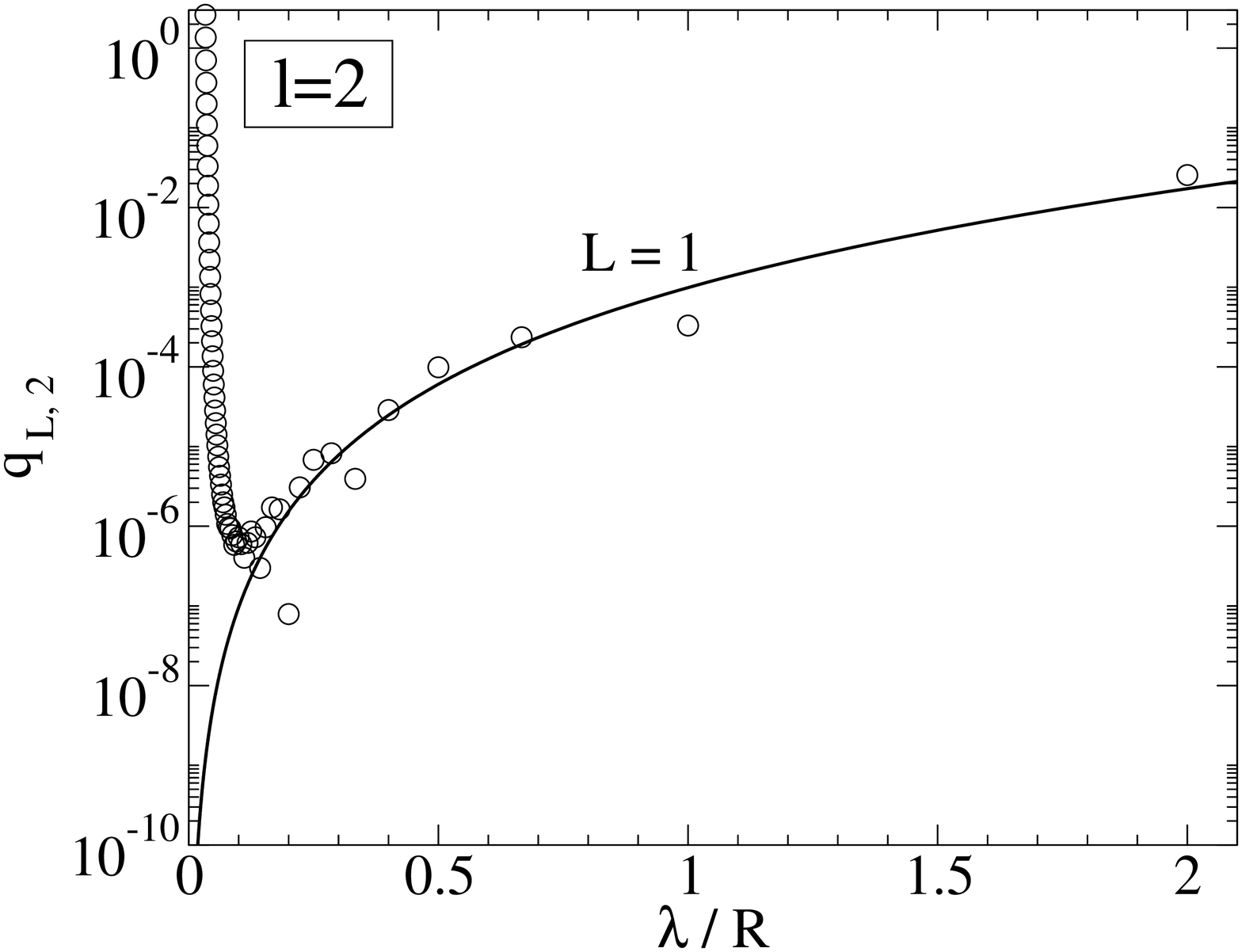} $\;$
\includegraphics[width=0.5\textwidth]{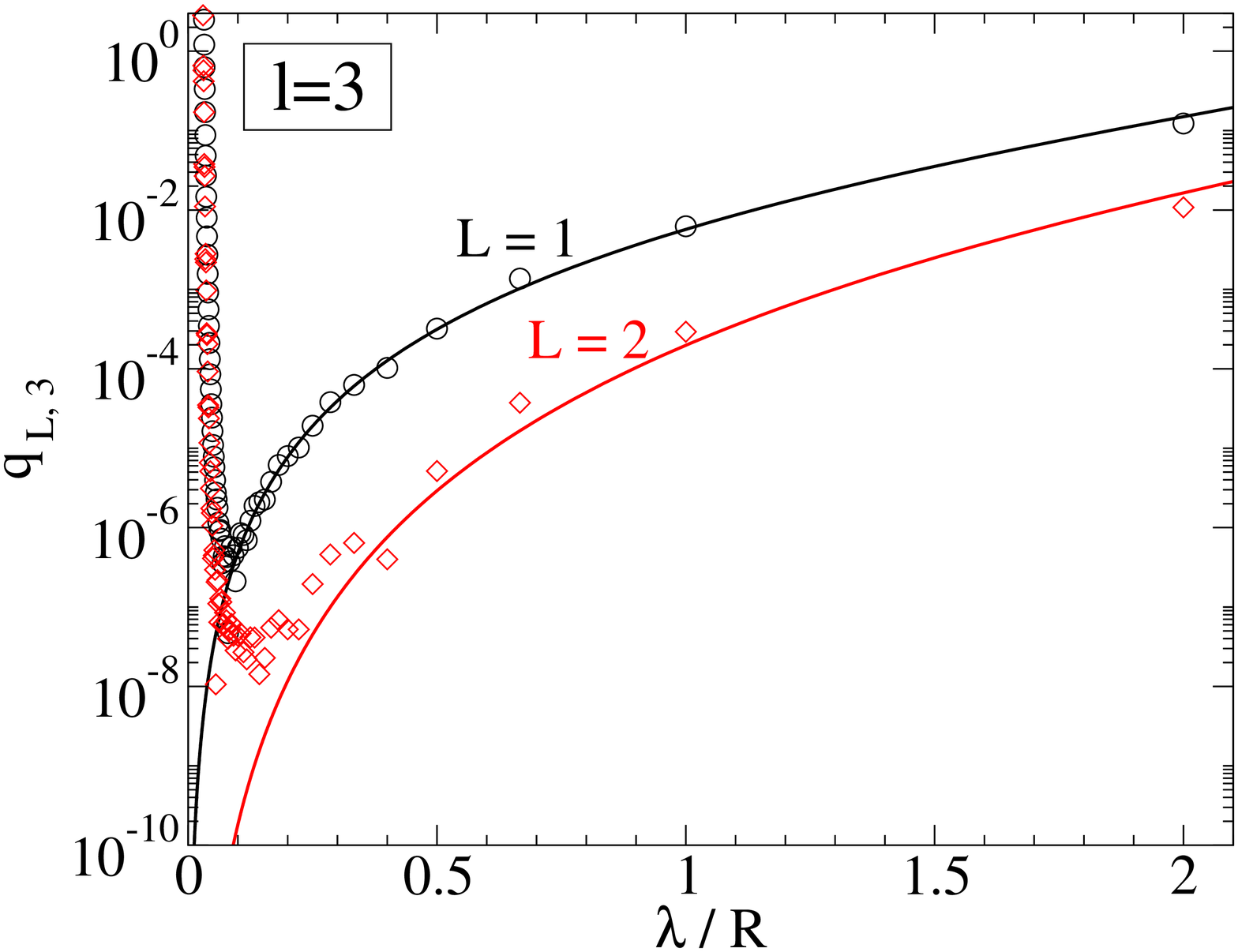}
\centerline{\includegraphics[width=0.5\textwidth]{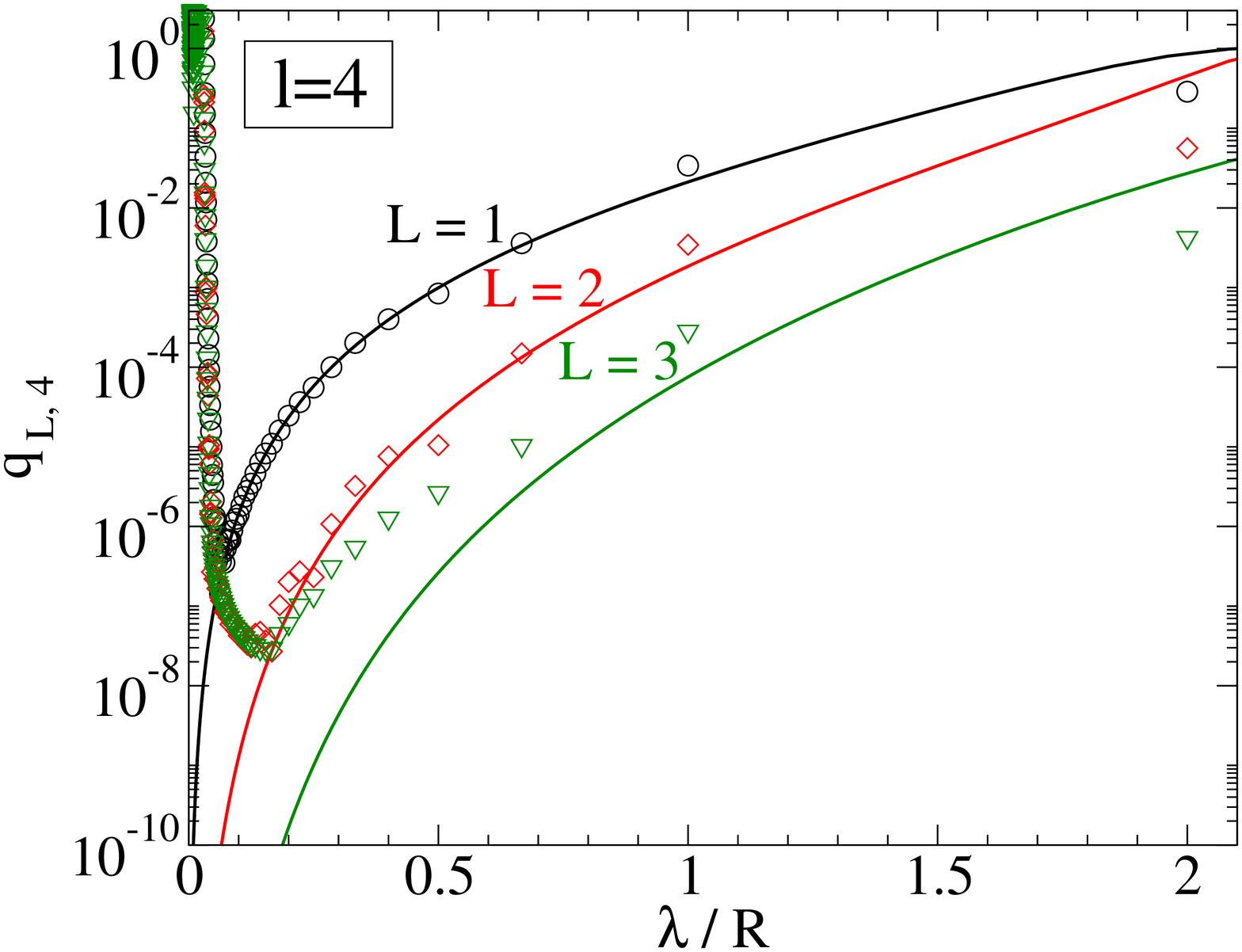}}
\caption{\label{f:ReflCoeff} 
  Reflection coefficients as a function of $\lambda/R$ (where $R$
  is the outer boundary radius and $\lambda$ is the wavelength) for
  odd-parity linearized solutions to the Einstein equations,
  with absorbing BCs \Hobc{L} of selected orders $L$.
  Curves denote coefficients $q_{L,\ell}$ calculated from equation 96 
  of \cite{Buchman2006}, and points denote coefficients computed
  from numerical evolutions of linearized-wave initial data. 
  Different panels correspond to solutions with different values of 
  the angular momentum number $\ell$.  
  Note that \Hobc{L=\ell} is perfectly absorbing and
  hence $q_{L,\ell}=0$.
}
\end{figure}

The predicted reflection coefficients in \cite{Buchman2006}
are given as functions of $2\pi R/ \lambda$, where $\lambda$ is 
the wavelength of the radiation and
$R$ is the radius of a spherical outer boundary.  Our numerical tests,
however, evolve wave packets and not monochromatic waves.  So to
obtain the reflection coefficients as a function of $\lambda$ for our
numerical evolutions, we proceed as follows. 
We measure the radiation reflected
off the outer boundary by taking the difference between our
numerically-evolved solution and the exact linearized solution at the
boundary, which we have plotted as a function of time in figures
\ref{f:L2DeltaPhi22K1Freezing}, \ref{f:L3DeltaPhi} and
\ref{f:L4DeltaPhi}.  
In order to minimize the nonlinear effects, 
the lowest amplitudes are used ($A=10^{-2.89}$ for $\ell=2$,
$A=10^{-3.5}$ for $\ell=3$ and $A=10^{-4}$ for $\ell=4$).
Taking the Fourier transform of any of the curves in these figures 
(prior to taking absolute values)
yields the difference as a function of frequency, from which
we can obtain the difference as a function of wavelength, $\Delta
\Phi(\lambda)$.  Similarly, we can obtain $\Phi(\lambda)$ by taking
the Fourier transform of the exact linearized solution $\Phi(t)$,
evaluated at the boundary.  Given these quantities, we define the
reflection coefficient at any particular wavelength $\lambda$ as
$\Delta \Phi(\lambda)/\Phi(\lambda)$.

The values of both the analytical and numerical
reflection coefficients are plotted as a function of $\lambda/R$ in
figure \ref{f:ReflCoeff}, for radiation with angular momentum numbers
$\ell=2,3,4$ and absorbing BCs \Hobc{L<\ell}. This demonstrates 
the importance of higher-order BCs when the numerical simulation
contains multipolar radiation. For example, assume the outer boundary
is located at twice the characteristic wavelength of the
simulation, so that $\lambda/R=0.5$. Then, for radiation with $\ell=3$
and freezing-$\Psi_0$ BCs (\Hobc{1}), the predicted errors due to
spurious reflections off the boundary are $~3\times 10^{-4}$. The use
of \Hobc{2} decreases these errors
$100$-fold, down to $~3\times 10^{-6}$. For multipolar radiation with
$\ell=4$, a factor of about $3600$ in accuracy is gained as one goes
from \Hobc{1} (where the error is $~9 \times 10^{-4}$)
to \Hobc{3} (where the error is $~ 2.5 \times
10^{-7}$).

Figure \ref{f:ReflCoeff} also shows that the predicted and numerical 
reflection coefficients agree well except for small values of $\lambda/R$
and small values of the reflection coefficient.
The discrepancies are likely to be caused by a combination of
different effects.
(i) We have computed the numerical reflection coefficients from a
fully nonlinear evolution whereas the predictions are only valid in
linearized theory. These nonlinear effects should become less
important as the amplitude of the wave is decreased.
(ii) Numerical roundoff error becomes important for small values of
the difference between the exact and numerical solution (visible in figures
\ref{f:L2DeltaPhi22K1Freezing}, \ref{f:L3DeltaPhi} and \ref{f:L4DeltaPhi}).
(iii) There is an accumulated error of spurious reflections passing through the
origin and reflecting again, which is not accounted for in the
theoretical reflection coefficients.  One can see this in the
secondary pulse at $t\approx 45$ in figures \ref{f:L2DeltaPhi22K1Freezing},
\ref{f:L3DeltaPhi} and \ref{f:L4DeltaPhi}.
(iv) For $\lambda \ell/(2 \pi R) \gtrsim 1$, the 
theoretical predictions for the reflection coefficients start to
break down because the ingoing solutions used for the calculation 
in \cite{Buchman2006} are not valid as the wave approaches $r=0$.


\section{Conclusions}
\label{s:Concl}

An algorithm for numerically implementing the higher-order absorbing
BCs \Hobc{L} for Einstein's equations presented in \cite{Buchman2006} has been
defined, tested, and shown to work.  Our method is based on a
reformulation of the BCs in terms of the gauge-invariant RWZ scalars.
This approach relies on the assumption that close to the outer
boundary, the Einstein equations can be linearized about a given
background spacetime, which in this paper is taken to be flat.  
It also requires a spherical outer boundary.
A key feature of our algorithm is the introduction of auxiliary
variables intrinsic to the boundary, a technique familiar from the
fields of, for example, computational electromagnetism and acoustics.

We have used a generalized harmonic formulation of the Einstein
equations \cite{Lindblom2006}; however, it should be straightforward
to adapt our algorithm to different formulations, with minor
modifications.  Similarly, our use of spectral numerical methods is
not an essential ingredient.  Our boundary algorithm can be added to
any existing numerical relativity code in a modular way: all that
needs to be done is to implement the auxiliary evolution system at the
boundary and provide for the necessary exchange of information with
the main evolution system.  In order to estimate the additional
computational cost in doing so, we note that the number of auxiliary
variables $w^{(\pm)}_{k,\ell m}$ that need to be evolved for the
absorbing BC \Hobc{L} (which is perfectly absorbing for all
multipoles $\ell \leqslant L$) is $2L(L+1)^2$. (The ranges are $1
\leqslant k \leqslant L$, $0 \leqslant \ell \leqslant L$, $-\ell
\leqslant m \leqslant \ell$.)  E.g.~for $L=3$, this amounts to $96$
functions (of time only), which is small compared to a typical number
of grid points in the interior domain.

We have implemented our method in the Caltech-Cornell SpEC code and
have tested it by evolving initial data with not only quadrupolar
($\ell=2$), but also higher multipolar ($\ell=3,4$) radiation,
imposing the hierarchy of BCs \Hobc{L} with $L \leq \ell$.  We
use linearized solutions of the Einstein equations
\cite{Teukolsky1982,Rinne2008c} as initial data, and evolve these in a
fully nonlinear code.  We demonstrate that with decreasing amplitude
of the radiation in the initial data, and perfectly absorbing BCs 
\Hobc{L=\ell}, the numerical solution decays at least quadratically to the
exact linearized solution, as expected.  For \Hobc{L<\ell},
the difference between the numerical and exact solutions
decays only linearly as the amplitude is decreased, indicating that
linear, spurious reflections are being introduced into the solution.
We have estimated the magnitude of these spurious reflections in our
numerical simulations and find good agreement with the theoretical
reflection coefficients derived in \cite{Buchman2006}, within the
range of validity of the two calculations. For multipolar radiation
with $\ell=3$ or $4$, we have seen that even without imposing a
perfectly absorbing BC \Hobc{L=\ell}, one can decrease the 
reflections off the outer boundary dramatically by increasing the
order $L$ of the BC \Hobc{L}. For instance, for an outer boundary
radius of twice the wavelength, $R=2\lambda$, and $\ell=3$, we achieve a
$100$-fold decrease in reflection by using \Hobc{2} rather than \Hobc{1}.

The most important application of our work is to the simulation of
isolated systems, such as coalescing binary black holes.  In these
simulations, the most sophisticated BC on the gravitational radiation
currently in use is the freezing-$\Psi_0$ BC
\cite{Bardeen2002,Kidder2005,Sarbach2005,Scheel2006,Rinne2007,
  Boyle2007,Scheel2008}, which corresponds to \Hobc{1}.
Note that the ratio of the wavelength to the outer boundary radius 
in the numerical tests presented in this
article is $\lambda/R\sim 0.1$; such a small value for $\lambda/R$
makes it easy to ensure that the BCs are consistent with the initial
data because the radiation vanishes at the outer boundary.  In
contrast, for binary black hole simulations, $\lambda/R$ is much
closer to unity, in particular during the inspiral phase. One can see
from the graphs in figure \ref{f:ReflCoeff} that errors due to
reflections with imperfectly absorbing BCs (including \Hobc{1})
are several orders of magnitude larger at, say, $\lambda/R = 0.5$
than they are at $\lambda/R = 0.1$.  
Furthermore, the majority of current binary black hole
simulations use far cruder BCs on the gravitational radiation than \Hobc{1}:
Typically, a sequence of adaptive mesh refinement \cite{Berger1984}
grids of decreasing resolution are used that extend out to larger and
larger radii  (see e.g.~\cite{Baker2006a,Diener2006,Husa2007,
LoustoZlochower2008,Hinder2008b}).
On the coarsest grid, the waves are no longer properly
resolved, and some {\it ad hoc} BC (such as a Sommerfeld BC) is imposed.  
It is clear that this approach will cause spurious reflections, and given
the results of \cite{Rinne2007}, we suspect that these will be
considerably larger than those caused by \Hobc{1}.

Inspiralling and merging equal mass non-spinning binary black holes
emit predominantly quadrupolar ($\ell=2$) radiation.  For a typical outer
boundary location used in numerical simulations of these events,
$R=2\lambda$, the errors due to spurious reflections 
of quadrupolar radiation
with \Hobc{1} are predicted in \cite{Buchman2006} to be $6
\times 10^{-5}$, which seems very small. However, these reflections
may interact with the evolution in such a way as to result in an
accumulating phase error.  For instance, the
runs in \cite{Boyle2007} showed phase errors of a few hundredths of 
a radian which were attributed to the outer boundary location.
Moreover, as the mass ratio $M_1/M_2$ deviates from unity, the energy 
emitted in non-quadrupolar modes rapidly increases: while for $M_1/M_2=1$,
less than 0.1 per cent of the energy is emitted with $\ell>2$, this fraction
increases to a few per cent for $M_1/M_2=2$
and exceeds $10$ per cent for the still fairly modest mass ratio
$M_1/M_2=4$ \cite{Berti-Cardoso-etal:2007,Berti2007b}.  
Because \Hobc{1} has larger reflection coefficients 
at a given value of $\lambda/R$ for $\ell>2$ modes than for 
$\ell=2$ modes (cf.~figure~\ref{f:ReflCoeff}), this BC will
not perform as well for 
unequal mass binary black hole simulations; in these cases,
the higher-order BCs presented in this paper will be even
more important than they are for equal mass simulations.
Finally, when considering the merit of the higher-order BCs on
simulations containing different modes $(\ell,m)$, one needs to
remember that the wavelength of a mode depends on the value of $m$.  
During the inspiral, modes with smaller $|m|$ have larger
wavelengths $\lambda$ (typically in proportion to $1/|m|$), 
and therefore the ratio of wavelength to boundary radius, 
$\lambda/R$, depends on $m$.  
From figure~\ref{f:ReflCoeff} we then deduce that modes
with the same $\ell$ but different $m$ will have different reflection
coefficients. 
For example, the $(2,1)$ mode has been shown to contribute
significantly to the unequal mass binary black hole inspiral 
radiation \cite{Berti-Cardoso-etal:2007}. 
Because its wavelength is longer than for the $(2,2)$ mode, 
the reflection coefficient for \Hobc{1} will be correspondingly higher.

For these reasons, we are confident that higher-order BCs will improve
the accuracy of long-term unequal mass binary black hole simulations.
Whether such an improvement is desirable will of course depend on
the precision requirements for the computed waveforms.  Parameter
estimation for LIGO requires phase errors of hundredths of a
radian~\cite{Lindblom2008}, and LISA will require yet more accurate
waveforms.

Our work can also be applied to the problem of Cauchy-perturbative
matching. In this approach, one matches solutions to the full
nonlinear Einstein equations on an interior spatial domain to
solutions of the linearized equations on an exterior domain extending
out to large radii.  These linearized equations can be represented for
instance by the gauge-invariant RWZ equations.  The question now
arises as to how one should impose outer BCs on the Einstein equations
in the interior with boundary data computed from the exterior
linearized solution.
Sections \ref{s:EinsteinBCs} and \ref{s:Reconstruction} contain a 
specific prescription for how this can be done.

In future work, we plan to generalize our algorithm by (i) including
first-order corrections in $M/R$ for the curvature and backscatter on
a Schwarzschild background spacetime (of mass $M$), (ii) allowing for
arbitrary coordinates of the background spacetime, and (iii) adapting
the algorithm to more general (i.e., not necessarily spherical)
boundary shapes. The generalized RWZ formalism \cite{Sarbach2001} upon
which our calculations are based and the absorbing BCs formulated in
\cite{Buchman2006,Buchman2007} allow for such generalizations.  In
more realistic situations where an exact solution is not at hand, one
can assess the quality of the BCs by comparing with a numerically
computed (fully nonlinear) reference solution on a large domain
\cite{Rinne2007}.  Ultimately, we plan to apply our implementation of
absorbing BCs to binary black hole simulations using the
Caltech-Cornell SpEC code.



\ack We thank James Bardeen, Edvin Deadman, Lee Lindblom, Richard
Matzner, Olivier Sarbach, Erik Schnetter, John Stewart and Manuel Tiglio for
insightful suggestions and discussions during the course of this work,
and Keith Matthews for use of and help with his ODE integration code.  
The numerical simulations presented here were performed using the 
Spectral Einstein Code (SpEC) developed at Caltech and Cornell 
primarily by Larry Kidder, Harald Pfeiffer and Mark Scheel.

This work was supported in part by grants to Caltech from the Sherman
Fairchild Foundation and the Brinson Foundation, by NSF grants
DMS-0553302, PHY-0601459, PHY-0652995, and by NASA grant NNG05GG52G.
LTB was also supported by grants NSF PHY 03 54842 and NASA NNG 04GL37G to
the University of Texas at Austin.  OR gratefully acknowledges funding
through a Research Fellowship at King's College Cambridge.


\appendix

\section{Conversion to spin-weighted harmonics}
\label{s:SpinWeighted}

The spin-weighted spherical harmonics \cite{NewmanPenrose1966} 
${}_s Y_{\ell m}(\theta,\phi)$ can be constructed from the 
standard spherical harmonics $Y_{\ell m}$ recursively using the relations
\begin{eqnarray}
  {}_0 Y_{\ell m} = Y_{\ell m},\\
  {}_{s+1} Y_{\ell m} = [(\ell-s)(\ell+s+1)]^{-1/2} \eth \, {}_s Y_{\ell m},\\
  {}_{s-1} Y_{\ell m} = -[(\ell+s)(\ell-s+1)]^{-1/2} \bar \eth \, {}_s Y_{\ell m}.
\end{eqnarray}
The operators $\eth$ and $\bar\eth$ are defined by
\begin{eqnarray}
  \eth {}_s Y_{\ell m} \equiv (-\partial_\theta 
  - \rmi \csc \theta \, \partial_\phi + s \cot \theta) {}_s Y_{\ell m},\\
  \bar \eth {}_s Y_{\ell m} \equiv (-\partial_\theta 
  + \rmi \csc \theta \, \partial_\phi - s \cot \theta) {}_s Y_{\ell m}.
\end{eqnarray}

Next we set up a tetrad $\be_\mu = (\mathbf{t, r, m, \bar m})$.
(Greek indices $\mu,\nu,\ldots$ from the middle of the alphabet denote
tetrad indices.)  In components with respect to spherical polar
coordinates $(t,r,\theta,\phi)$,
\begin{equation}
  t_\alpha = (1,0,0,0), \quad r_\alpha = (0,1,0,0), \quad
  m_\alpha = \frac{r}{\sqrt{2}} (0,0,1,\rmi\sin\theta),
\end{equation}
and $\mathbf{\bar m}$ is the complex conjugate of $\mathbf{m}$.
An arbitrary rank-2 tensor $\mathbf{T}$ can be expanded as
\begin{equation}
  \label{e:SpinWeightHarmDecomp}
  \mathbf{T} = \sum_{\ell,m,\mu,\nu} T^{\be_\mu \be_\nu}_{\ell m} \; 
  {}_{s(\be_\mu,\be_\nu)} Y_{\ell m} \; \be_\mu \otimes \be_\nu.
\end{equation}
Here the spin weight is defined by 
$s(\be_\mu, \be_\nu) = s(\be_\mu) + s(\be_\nu)$
where $s(\mathbf{m}) = -1$, $s(\mathbf{\bar m}) = +1$, and 
$s(\mathbf{t}) = s(\mathbf{r}) = 0$.
The SpEC code provides routines that compute the expansion
coefficients $T^{\be_\mu \be_\nu}_{\ell m}$ in \eref{e:SpinWeightHarmDecomp}
for a given tensor $\mathbf{T}$.

The basis harmonics defined in \eref{e:YSharmonics} can be written as
\begin{eqnarray}
  Y_{A,\ell m} = -\frac{\sqrt{\ell(\ell+1)}}{\sqrt{2}\, r} 
    ({}_1Y_{\ell m} \bar m_A - {}_{-1}Y_{\ell m} m_A),\\
  S_{A,\ell m} = -\frac{i \sqrt{\ell(\ell+1)}}{\sqrt{2}\, r} 
    ({}_1Y_{\ell m} \bar m_A + {}_{-1}Y_{\ell m} m_A),\\
  Y_{AB,\ell m} = \frac{\sqrt{\lambda\ell(\ell+1)}}{2r^2}
    ({}_2Y_{\ell m} \bar m_A \bar m_B + {}_{-2}Y_{\ell m} m_A m_B),\\
  S_{AB,\ell m} = \frac{i\sqrt{\lambda\ell(\ell+1)}}{2r^2}
    ({}_2Y_{\ell m} \bar m_A \bar m_B - {}_{-2}Y_{\ell m} m_A m_B).
\end{eqnarray}
(Recall that $\lambda = (\ell-1)(\ell+2)$.)
We also have $\hat g_{AB} = 2 r^{-2} m_{(A} \bar m_{B)}$.
Using these relations it is straightforward to express the amplitudes
of the perturbation \eref{e:OddDecomp1}--\eref{e:OddDecomp2} and
\eref{e:EvenDecomp1}--\eref{e:EvenDecomp3}
in terms of the expansion coefficients $\delta g^{\be_\mu \be_\nu}_{\ell m}$,
\begin{eqnarray}
  h_{t,\ell m} = \frac{\rmi r}{\sqrt{2\ell(\ell+1)}} 
    (\delta g^\mathbf{t\bar m}_{\ell m} + \delta g^\mathbf{tm}_{\ell m}),\\
  h_{r,\ell m} = \frac{\rmi r}{\sqrt{2\ell(\ell+1)}} 
    (\delta g^\mathbf{r\bar m}_{\ell m} + \delta g^\mathbf{rm}_{\ell m}),\\
  k_{\ell m} = \frac{-\rmi r^2}{2\sqrt{\lambda\ell(\ell+1)}}
    (\delta g^\mathbf{\bar m \bar m}_{\ell m} - \delta g^\mathbf{mm}_{\ell m}),\\
  H_{tt,\ell m} = \delta g^\mathbf{tt}_{\ell m},\\
  H_{tr,\ell m} = \delta g^\mathbf{tr}_{\ell m},\\
  H_{rr,\ell m} = \delta g^\mathbf{rr}_{\ell m},\\
  Q_{t,\ell m} = \frac{-r}{\sqrt{2\ell(\ell+1)}}
    (\delta g^\mathbf{t\bar m}_{\ell m} - \delta g^\mathbf{tm}_{\ell m}),\\
  Q_{r,\ell m} = \frac{-r}{\sqrt{2\ell(\ell+1)}}
    (\delta g^\mathbf{r\bar m}_{\ell m} - \delta g^\mathbf{rm}_{\ell m}),\\
  K_{\ell m} = \delta g^\mathbf{m\bar m}_{\ell m},\\
  G_{\ell m} = \frac{1}{\sqrt{\lambda\ell(\ell+1)}}
    (\delta g^\mathbf{\bar m \bar m}_{\ell m} + \delta g^\mathbf{mm}_{\ell m}).
\end{eqnarray}


\section*{References}

\bibliographystyle{oriop}
\bibliography{../References/References}

\end{document}